\documentclass[prd, 11pt,nofootinbib, superscriptaddress, preprintnumbers,floatfix]{revtex4}
\usepackage{amssymb,amsmath}
\usepackage{graphicx}
\usepackage{multirow}
\usepackage{bm}
\usepackage{comment}
\usepackage{color}
\usepackage{slashed}

\begin{document}
\newcommand{\ignore}[1]{}
\def\mc#1{{\mathcal #1}}
\newcommand{\be}{\begin{equation}}
\newcommand{\ee}{\end{equation}}
\newcommand{\ba}{\begin{eqnarray}}
\newcommand{\ea}{\end{eqnarray}}
\newcommand{\order}[1]{\mathcal{O}(#1)}
\newcommand{\cir}[1]{\mathring{#1}}
\graphicspath{{./plots/}}

\title{Nucleon isovector scalar charge from overlap fermions}

\author{Liuming Liu}
\email{liuming@impcas.ac.cn}
\affiliation{Institute of Modern Physics, Chinese Academy of Sciences, Lanzhou, 730000, China}
\affiliation{University of Chinese Academy of Sciences, Beijing 100049, China. }

\author{Ting Chen}
\affiliation{School of Physics, Peking University, Beijing 100871, China}

\author{Terrence Draper}
\affiliation{Department of Physics and Astronomy, University of Kentucky, Lexington, KY 40506, USA}

\author{Jian~Liang}
\email{jianliang@scnu.edu.cn}
\affiliation{Guangdong Provincial Key Laboratory of Nuclear Science, Institute of Quantum Matter, South China Normal University, Guangzhou 510006, China}
\affiliation{Guangdong-Hong Kong Joint Laboratory of Quantum Matter, Southern Nuclear Science Computing Center, South China Normal University, Guangzhou 510006, China}

\author{Keh-Fei~Liu}
\email{liu@g.uky.edu}
\affiliation{Department of Physics and Astronomy, University of Kentucky, Lexington, KY 40506, USA}

\author{Geng Wang}
\affiliation{Department of Physics and Astronomy, University of Kentucky, Lexington, KY 40506, USA}

\author{Yi-Bo Yang}
\email{ybyang@mail.itp.ac.cn}
\affiliation{CAS Key Laboratory of Theoretical Physics, Institute of Theoretical Physics, Chinese Academy of Sciences, Beijing 100190, China}
\affiliation{School of Fundamental Physics and Mathematical Sciences, Hangzhou Institute for Advanced Study, UCAS, Hangzhou 310024, China}
\affiliation{International Centre for Theoretical Physics Asia-Pacific, Beijing/Hangzhou, China}

\collaboration{$\chi$QCD Collaboration }


\begin{abstract}
We calculate the nucleon isovector scalar charge in lattice QCD using overlap fermions on five ensembles of gauge configurations generated by the RBC/UKQCD collaboration using the domain-wall quark action with $2+1$ dynamical flavors. The five ensembles cover five pion masses, $m_\pi \approx$ 139, 171, 302, 337 and 371 MeV, and four lattice spacings, $a \approx $ 0.06, 0.08, 0.11 and 0.14 fm. Three to six valence quark masses are computed on each ensemble to investigate the pion mass dependence. The extrapolation to the physical pion mass, continuum and infinite volume limits is obtained by a global fit of all data to a formula originated from partially quenched chiral perturbation theory. The excited-states contamination is carefully analyzed with 3--5 sink-source separations and multi-state fits. Our final result, in the $\overline{\text{MS}}$ scheme at 2 GeV, is $g_{S}^{u-d}= 0.94 (10)_{stat}(8)_{sys}$, where the first error is the statistical error and the second is the systematic error. 
\end{abstract}

\maketitle

\newpage

\section{Introduction}

The nucleon scalar charge is a fundamental quantity in understanding the internal structure of nucleons and more importantly it is related to the search for new physics beyond the Standard Model (BSM). Together with the tensor charge, it probes novel scalar and tensor interactions at the TeV scale~\cite{Bhattacharya:2011qm}. The nucleon scalar charge is also an important input in the direct search for dark matter~\cite{Bottino:1999ei, Bottino:2008mf,Ellis:2008hf, Giedt:2009mr}. There are numerous ongoing or planned experiments targeted at searching for scalar and tensor interactions~\cite{Wilburm:2009, Pocanic:2008pu,Bischer:2018zcz,Akimov:2017ade, alonso2017isodarkamlanda, Akerib:2015cja, Beda:2012zz}. High precision experimental measurements would require precise input of the scalar/tensor charge to put stringent bounds on the existence of new physics. Unlike the axial charge, the scalar and tensor charges are not well known in experiments. Lattice QCD provides a first-principles non-perturbative formulation for numerical calculation of the fundamental quantities of the QCD theory with controlled uncertainties. In lattice QCD, the nucleon charges are extracted from the matrix elements of the local quark bilinear operators within the nucleon state. For the isovector charges, only the connected insertions are involved and thus are straightforward to compute. As an extension of our previous work on the isovector axial and tensor charges~\cite{Liang:2018pis, Liang:2016fgy, Horkel:2020hpi}, we compute the isovector scalar charge in this work. 

We use a mixed-action approach with overlap fermions in the valence sector and domain-wall configurations. Since both domain-wall and overlap fermions are chiral fermions, the non-perturbative renormalization via chiral Ward identities or RI/MOM scheme can be implemented relatively easily and the systematic uncertainty due to the use of an action explicitly breaking chiral symmetry can be avoided at finite lattice spacing. The multi-mass algorithm for overlap fermions allows us to calculate quark propagators for many different quark masses without much additional cost. Five ensembles covering five pion masses including one at the physical value, four lattice spacings in the range 0.06 fm - 0.14 fm and five volumes are used in this work, and 3-6 valence pion masses are computed for each ensemble. This enables us to make a reliable extrapolation to the physical pion mass, continuum limit and infinite-volume limit. Excited-states contamination is a main source of systematic uncertainty in lattice calculations of nucleon matrix elements. In order to investigate the excited-states contamination, we explicitly fit up to three states in the correlation functions with 3--5 source-sink separations. 

This paper is organized as follows. The numerical details about the lattice setup and the computation of correlation functions are presented in Sec.~\ref{sec:NumericalDetails}. The analysis of the correlation functions, in particular the investigation of the exited-state contamination,  is presented in Sec.~\ref{sec:CorrAnalysis}. In Sec.~\ref{sec:Renormalization}, we describe the renormalization procedure. In Sec.~\ref{sec:Results}, we present the renormalized values of the scalar charge and perform the extrapolation to the physical point. A summary is given in Sec.~\ref{sec:Summary}.

\section{Numerical details}
\label{sec:NumericalDetails}

\subsection{Lattice setup}
\label{subsec:LatticeSetup} 
The results presented in the paper are based on the gauge configurations generated by the RBC/UKQCD collaboration with 2+1 flavor domain-wall fermions~\cite{Aoki:2010dy, Arthur:2012yc, Blum:2014tka}. The relevant parameters of the five ensembles used in this work are collected in Table~\ref{Table:ensembles}. The gauge action is the Iwasaki+DSDR action~\cite{Vranas:2006zk, Fukaya:2006vs, Renfrew:2009wu} for the ensemble 32ID and the Iwasaki action~\cite{Iwasaki:2011np} for the rest of the ensembles. See Refs.~\cite{Aoki:2010dy, Arthur:2012yc, Blum:2014tka} for more details about the ensembles.

For the valence quark, we use the overlap fermion action~\cite{Neuberger:1997fp}. The overlap Dirac operator is defined as
\be
D_{ov} = 1 + \gamma_5 \epsilon(\gamma_5 D_w(\rho)),
\ee
where $\epsilon$ is the matrix sign function and $D_w$ is the Wilson Dirac operator with a negative mass parameter $-\rho = 1/2\kappa -4$. We set $\kappa = 0.2$ in our calculation which corresponds to $\rho=1.5$. The massive overlap Dirac operator is 
\be
D_m = \rho D_{ov}(\rho) + m \left(1-\frac{D_{ov} (\rho)}{2}\right).
\ee
To accommodate the $SU(3)$ chiral transformation, it is usually convenient to use the chirally regulated field $\hat{\psi} = (1-\frac{1}{2}D_{ov})\psi$ in lieu of $\psi$ in the interpolating field and currents. This is equivalent to leaving the currents unchanged and adopting the effective propagator~\cite{Chiu:1998eu, Liu:2002qu}
\be
G \equiv \left(1 - \frac{D_{ov}}{2}\right) D_m^{-1} = \frac{1}{D_c + m},
\ee
where $D_c = \frac{ \rho D_{ov}} {1-D_{ov}/2}$ is exactly chiral~\cite{Chiu:1998gp}, i.e., $\{\gamma_5, D_c\} = 0$. 

\begin{table}[t!]
\centering
\begin{tabular*}{\textwidth}{@{\extracolsep{\fill}}lcccccccr}
\hline
\hline
Ens. ID & $L^3 \times T$    &$a$(fm) & $m_l $ & $m_s $ & $m_\pi^{sea}$ (MeV) &$m_\pi^{val}$ (MeV) & $t_{sink}/a$ & $N_{conf}$ \\
\hline
32Ifine & $32^3 \times 64$ &0.0626 &0.0047 &0.0186 &371 &344, 383, 420 &14,16,18 &459 \\
\hline
32I & $32^3 \times 64$  &0.0828 &0.004 &0.03 &302 &294, 316, 353, 409 &12,14,15 &309 \\
\hline
24I &  $24^3 \times 64$  &0.1105  &0.005 &0.04 &340 &282, 321, 348, 389 &8,10,12 &203 \\
\hline
48I & $48^3 \times 96$  &0.1141 &0.00078 &0.0362 &139  &149, 181, 207, 267, 331, 372 &8,10,12 &81 \\
\hline
32ID & $32^3 \times 64$  &0.1432 &0.001 &0.045 &171 &174, 233, 262, 288, 327 &7,8,9,10,11 &200 \\
\hline
\hline
\end{tabular*}
\caption{Parameters of the ensembles. The labeling of the ensembles follows the notations in Ref.~\cite{Aoki:2010dy, Arthur:2012yc, Blum:2014tka}. The lattice volume $L^3 \times T$, lattice spacing $a$, bare light(strange) quark mass $m_l$($m_s$), unitary pion mass $m_{\pi}^{sea}$ and the number of configurations $N_{conf}$ are listed. For each ensemble, quark propagators are computed with multiple quark masses corresponding to the valence pion masses $m_{\pi}^{val}$. $t_{sink}$ is the source-sink separation of the correlation functions. See below for further explanation.}
\label{Table:ensembles}
\end{table} 

\subsection{Computation of the correlation functions}
\label{subsec:ComputeCorrFunc}
The nucleon isovector scalar charge is defined through the nucleon matrix element
\begin{equation}
\langle N|\bar{u}u - \bar{d}d |N \rangle = g_S \bar{u}_N u_N,
\end{equation}
where $u_N$ is the nucleon spinor at zero momentum with the normalization $\sum_{s} u_N(s) \bar{u}_N(s) =\frac{1+\gamma_4}{2}$, and $u$ and $d$ are up and down quark fields. This charge can be obtained from the ratio of the three-point function to the two-point function
\begin{equation}
R(t_{sink}, t)  = \frac{C_{3pt}(t_{sink}, t)}{C_{2pt}(t_{sink})} = \frac{\Gamma^e_{\alpha \beta} \langle 0| \sum_{\vec{y}} \chi_\alpha(t_{sink}, \vec{y})\, \mathcal{O}_S (t)\, \sum_{\vec{x}\in \mathcal{G}} \bar{\chi}_\beta(0, \vec{x}) |0 \rangle} { \Gamma^e_{\alpha \beta} \langle 0| \sum_{\vec{y}}\chi_\alpha(t_{sink}, \vec{y}) \sum_{\vec{x}\in \mathcal{G}} \bar{\chi}_\beta(0, \vec{x}) |0 \rangle},
\end{equation}
where $\chi$ is the proton interpolating operator $\chi = \epsilon^{abc}[{u^a}^T C\gamma_5 d^b] u^c$ and $\Gamma^e = \frac{1}{2}(1+\gamma_4)$ is the positive parity projector for the nucleon propagating in the forward direction. The scalar current is $\mathcal{O}_S(t) = \sum_{\vec{x}} \bar{q}(t, \vec{x}) q(t, \vec{x})$ with $q$ representing the $u/d$ quark fields. The current insertion time $t$ varies between the source and sink time locations. 
Smeared grid sources with Z3 noise~\cite{Dong:1993pk, Li:2010pw, Gong:2013vja} are used to compute the high-mode part of the
quark propagators, while the low-mode part is constructed exactly using the eigenvectors for each point of
the source grid $\cal G$. The correlation functions are calculated by combining the high- and low-mode parts using the low-mode substitution technique~\cite{Li:2010pw}, which helps to reduce the corresponding statistical uncertainties significantly.
To suppress the excited-states contamination, Gaussian smearing is applied to all quarks at sink and source. 
The calculation strategy is the same as in our previous works, e.g., ~\cite{Yang:2015uis, Liang:2018pis, Yang:2018nqn}, and please refer to these references for further details.

One of the advantages of overlap fermions is that one can compute the quark propagators with multiple quark masses at a small additional cost compared to the cost for the lightest quark mass. We employ 3--6 quark masses for each of the ensembles~\cite{Ying:1996je}. The corresponding valence pion masses $m_{\pi}^{val}$ are listed in Table~\ref{Table:ensembles}. 

In order to reduce the excited-states contamination, the source-sink separation $t_{sink}$ has to be large enough and the insertion time $t$ should be far away from both the source and the sink. For each ensemble, we calculate the correlation functions for a number of values of $t_{sink}$ and at least two of them are above 1 fm. The values of $t_{sink}$ are listed in Table~\ref{Table:ensembles}. The data at all $t_{sink}$ will be fitted simultaneously to extract the scalar charge. 

\begin{figure}
\includegraphics[width =0.4 \textwidth]{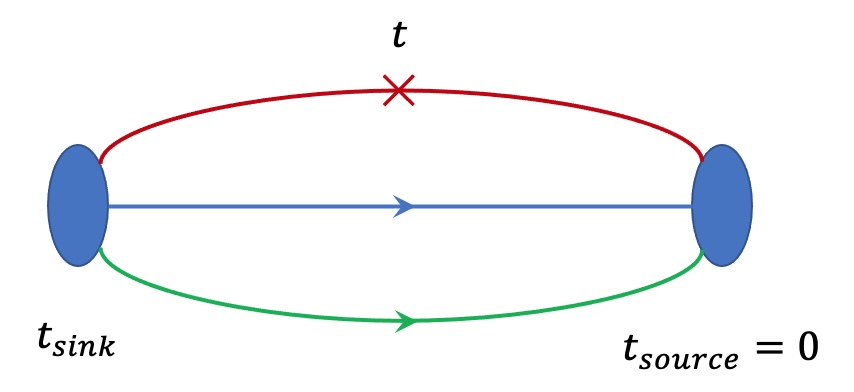}
\caption{Illustration of the connected insertion.}
\label{Fig:ConnectedInsertion}
\end{figure}

In this work we are interested in the isovector scalar charge, which means only the connected insertion, as illustrated in Fig.~\ref{Fig:ConnectedInsertion}, needs to be considered when calculating the three-point functions. 

\section{Analysis of the correlation functions}
\label{sec:CorrAnalysis}
Nucleon charges are given by the matrix elements of the quark bilinear operators between the nucleon ground state. The nucleon interpolating operator used in the calculation contains contributions from excited states. We use the so-called two-state fit, in which the contribution of the first excited state is taken into account, to extract the desired matrix element. Keeping the ground state and the first excited state in the spectral decomposition, the ratio of the three-point function to the two-point function can be written as
 \be
R(t_{sink}, t) = \frac{C_{3pt}(t_{sink}, t)}{C_{2pt}(t_{sink})} =  g_S + C_1 e^{-\Delta M_1 t_{sink}} + C_2 e^{-\Delta M_1 (t_{sink}-t)} + C_3 e^{-\Delta M_1 t},
\label{Eq:2statefit}
\ee
where $g_S$ is the scalar charge and $\Delta M_1$ is the mass difference between the first excited state and the ground state. We use this empirical form to describe the contamination from the excited states. In practice, $\Delta M_1$ can be considered as the effective weighted average of the mass differences between several excited states and the ground state. It is usually higher than the mass difference between the first excited state and the ground state. In Eq.~\ref{Eq:2statefit}, the higher powers of $e^{-\Delta M_1 t_{sink} }$ are dropped since they are negligibly small. We compute the ratio for the $u$ and $d$ flavors separately, denoted by $R^u(t_{sink}, t)$ and $R^d(t_{sink}, t)$, respectively, and perform a joint fit of $R^u(t_{sink}, t)$ and $R^d(t_{sink}, t)$ at all $t_{sink}$ simultaneously using the above formula. The parameter $\Delta M_1$ is common for $R^u(t, t_s)$ and $R^d(t, t_s)$. Thus there are 7 parameters to be determined in the fit: $g_S^u$, $g_S^d$, $C_1^u$, $C_1^d$, $C_2^u$, $C_2^d$, $C_3^u$, $C_3^d$ and $\Delta M$. The isovector scalar charge is then given by $g_S^{u-d} = g_S^u - g_S^d$. 

The data points with $t$ close to the sink or source suffer large excited-states contamination. Those data points should not be included in the fits. In our fits, three points at the source and sink ends were dropped for the ensembles 24I, 32I, 32ID, and 48I, while four points were dropped for the ensemble 32Ifine. This guarantees that the distance between the inserted current and the sink/source is around or larger than 0.25 fm, and also that the $\chi^2$ values of all fits are in reasonable range. 

In order to check the contributions from the higher excited states, we performed a three-state fit which retains three states in the spectral decomposition. The ratio of the three-point function to the two-point function takes the form
\be
R(t_{sink}, t) = g_S + C_1 e^{-\Delta M_1 t_{sink}} + C_2 e^{-\Delta M_1 (t_{sink}-t)} + C_3 e^{-\Delta M_1 t}  + C_4 e^{-\Delta M_2 t_{sink}} + C_5 e^{-\Delta M_2 (t_{sink}-t)} + C_6 e^{-\Delta M_2 t},
\label{Eq:3statefit}
\ee
where $\Delta M_2$ is the mass difference between the second excited state and the ground state. The terms arsing from the transition matrix elements between the first and second excited states are dropped in the above equation. They are found to be insignificant in the fits and thus are ignored for better stability in the fits. As for the two-state fit case, we perform a joint fit of $R^u(t_{sink}, t)$ and $R^d(t_{sink}, t)$ at all $t_{sink}$ simultaneously, keeping the parameters $\Delta M_1$ and $\Delta M_2$ common for $R^u(t_{sink}, t)$ and $R^d(t_{sink}, t)$. 

\begin{figure}
\includegraphics[width =0.33 \textwidth]{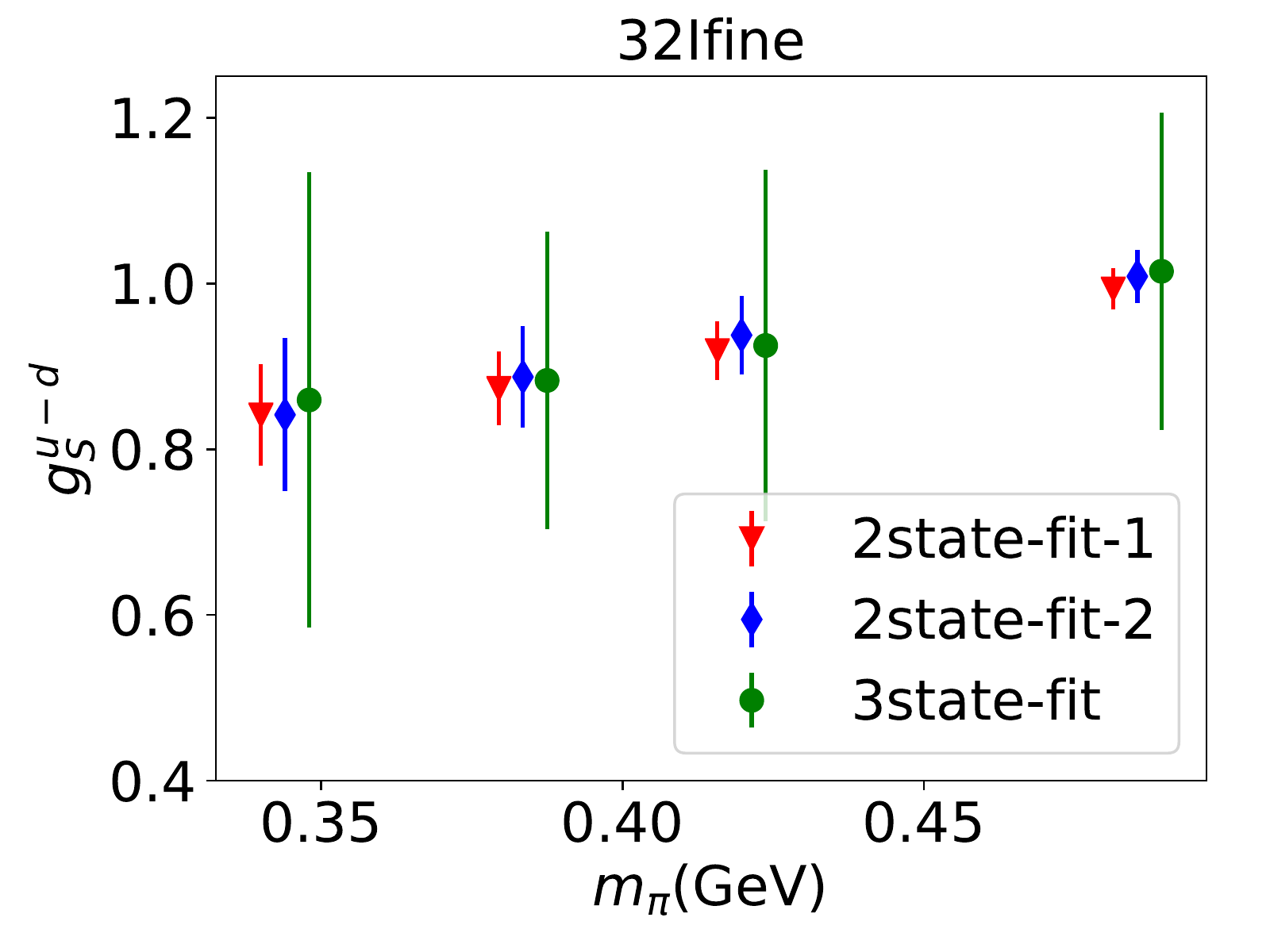}\includegraphics[width =0.33 \textwidth]{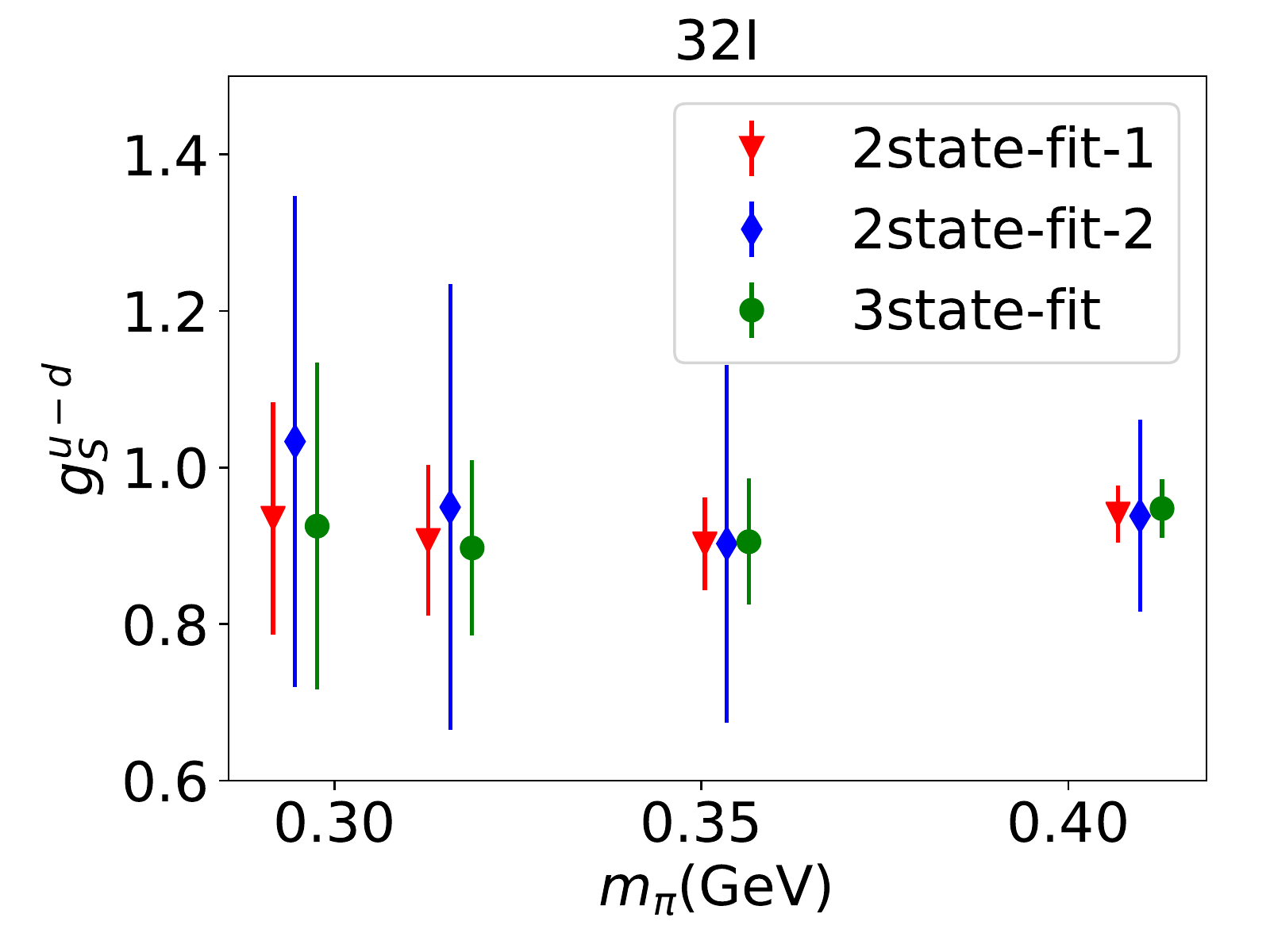}\includegraphics[width =0.33 \textwidth]{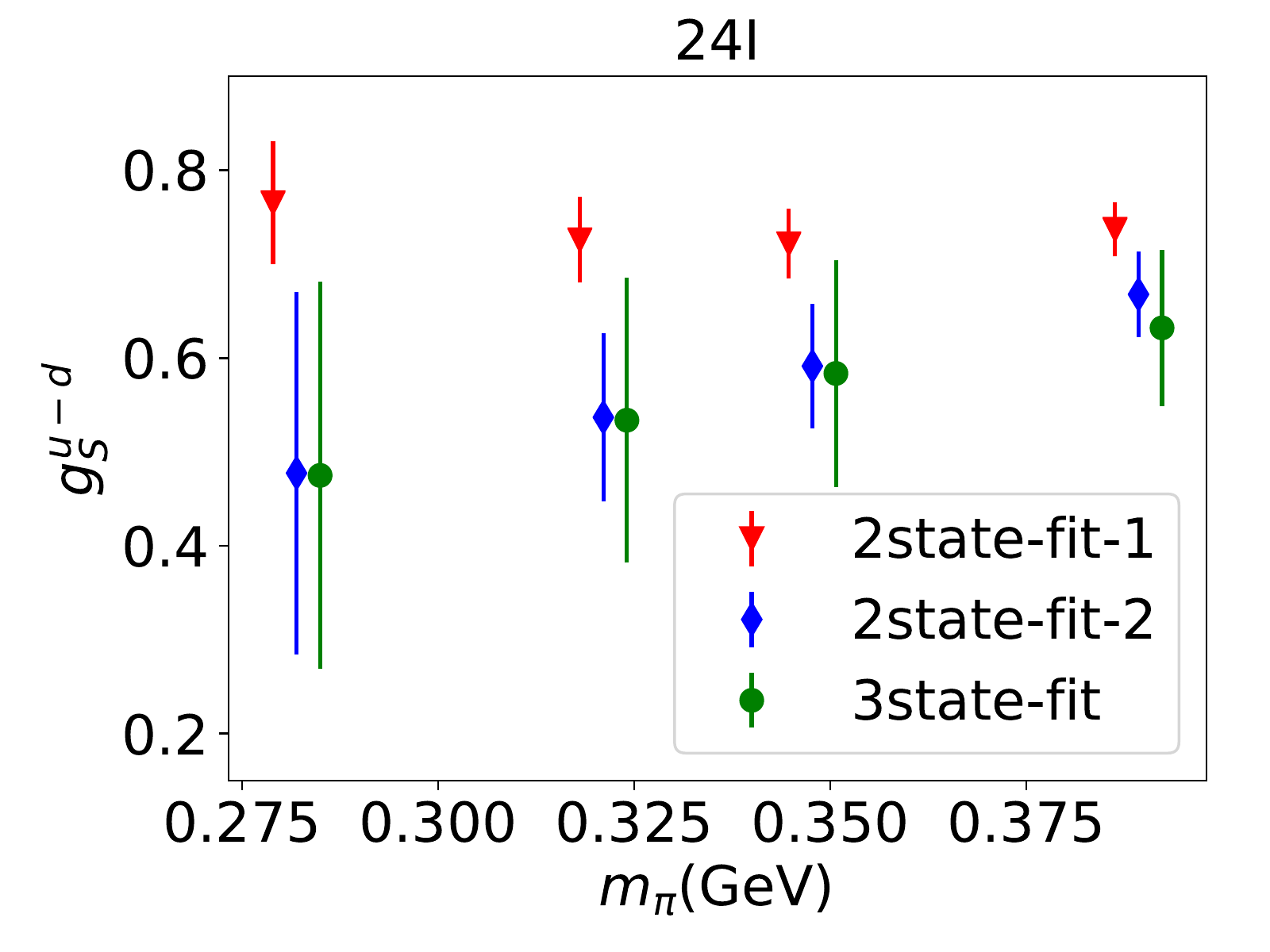}
\includegraphics[width =0.33 \textwidth]{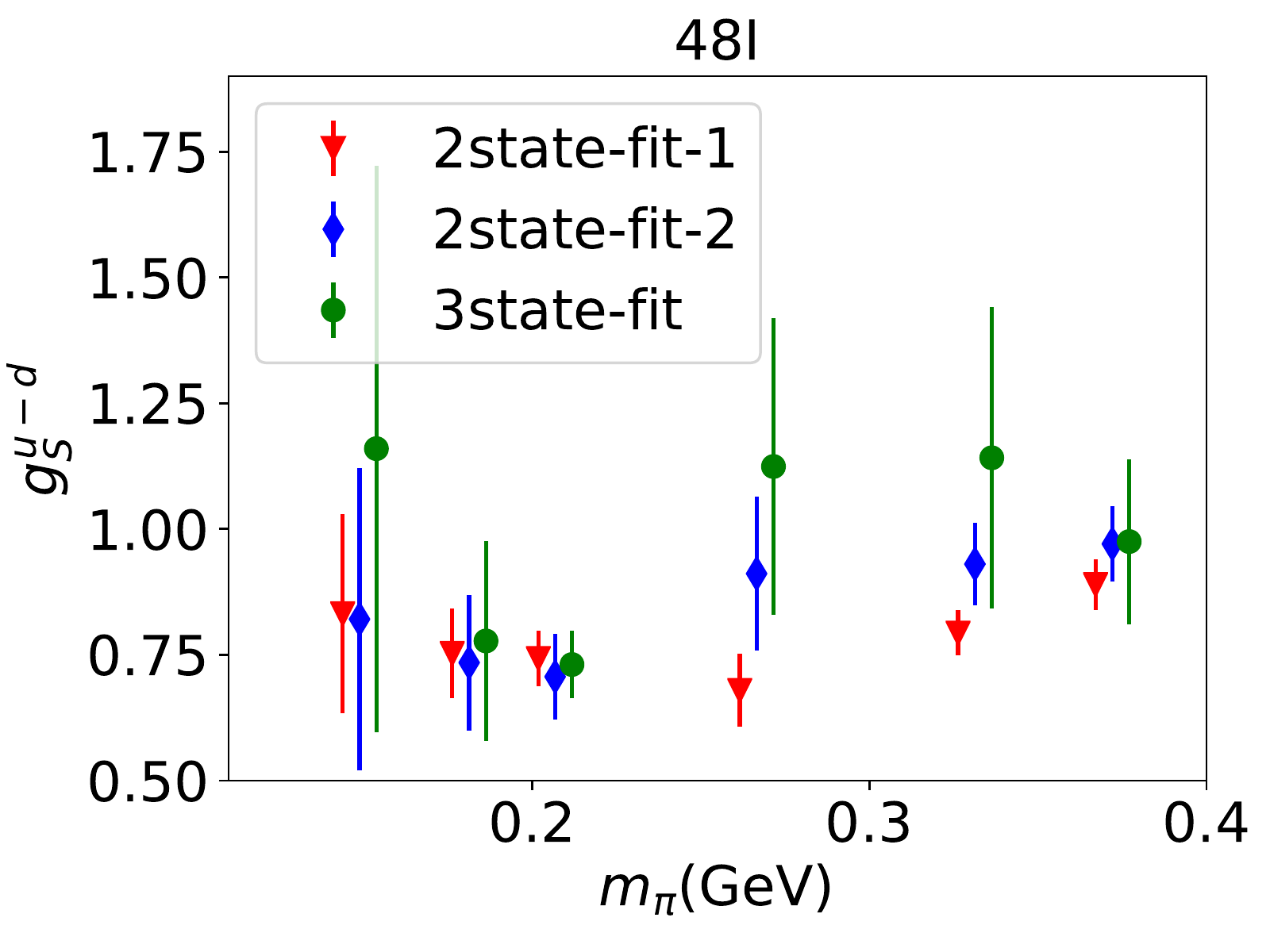}\includegraphics[width =0.33 \textwidth]{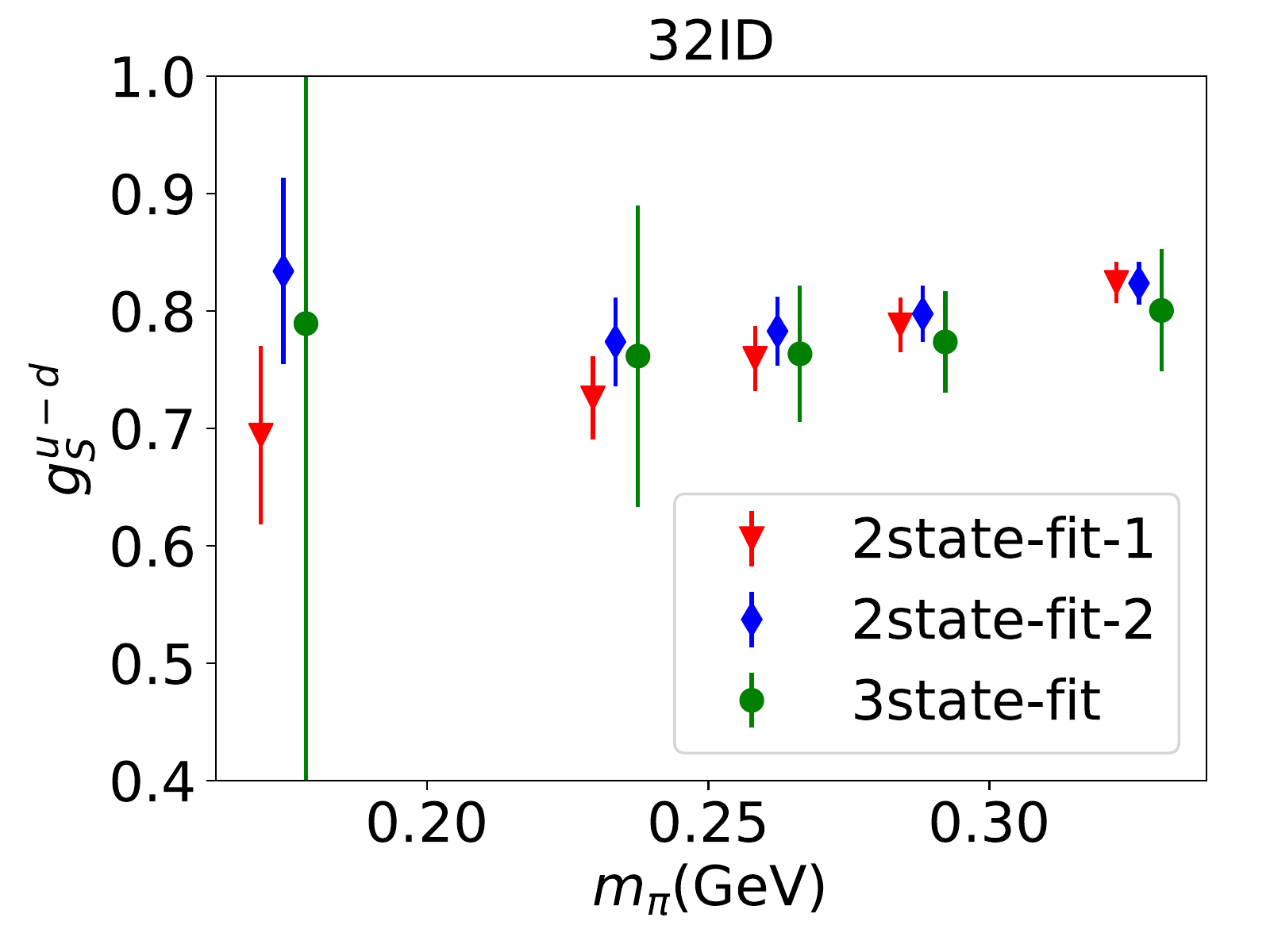}
\caption{Comparing $g_S$ values from two-state fits and three-state fits.}
\label{Fig:compare_gs}
\end{figure}

Comparing the results of the two- and three-state fits, we found some discrepancies in ensembles 24I and 48I. Note that for these two ensembles the smallest value of $t_{sink}$ is $\sim 0.88$ fm, which is rather small and may suffer from large excited-states contamination. We drop the data points with the smallest value of $t_{sink}$ and redo the two-state fit for all ensembles. In Fig.~\ref{Fig:compare_gs}, we compare the results of the three types of fit: 1) two-state fit with all data points, labeled as ``2state-fit-1" in the figure and the following text; 2) two-state fit excluding the data points with the smallest $t_{sink}$, labeled as ``2state-fit-2"; 3) three-state fit with all data points, labeled as ``3state-fit". One can see that the results of the three types of fits agree with each other very well for the ensembles 32I, 32ID and 32Ifine, while for the ensembles 24I and 48I, ``2state-fit-2" agrees better with ``3state-fit" than the ``2state-fit-1" does. Therefore, we take the results of ``2state-fit-2" for the ensembles 24I and 48I and ``2state-fit-1" results for the ensembles 32I, 32ID and 32Ifine as our two-state fit results. The difference between the two-state fit and the three-state fit results will be taken as an estimation of the systematic uncertainty due to excited-states contamination. The two- and three-state fit results of unrenormalized $g_S^{u-d}$ for all ensembles are collected in Table~\ref{Table:gs_results}.

\begin{figure}
\includegraphics[width =0.12 \textwidth]{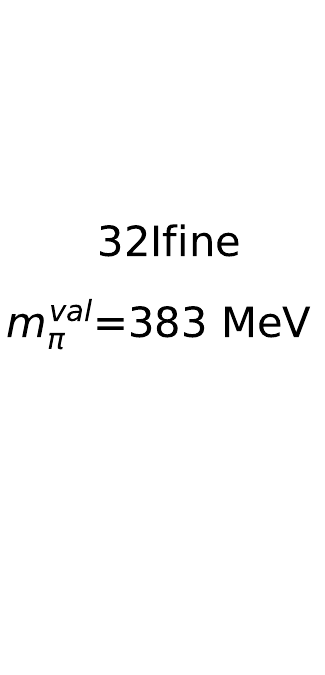}\includegraphics[width =0.44 \textwidth]{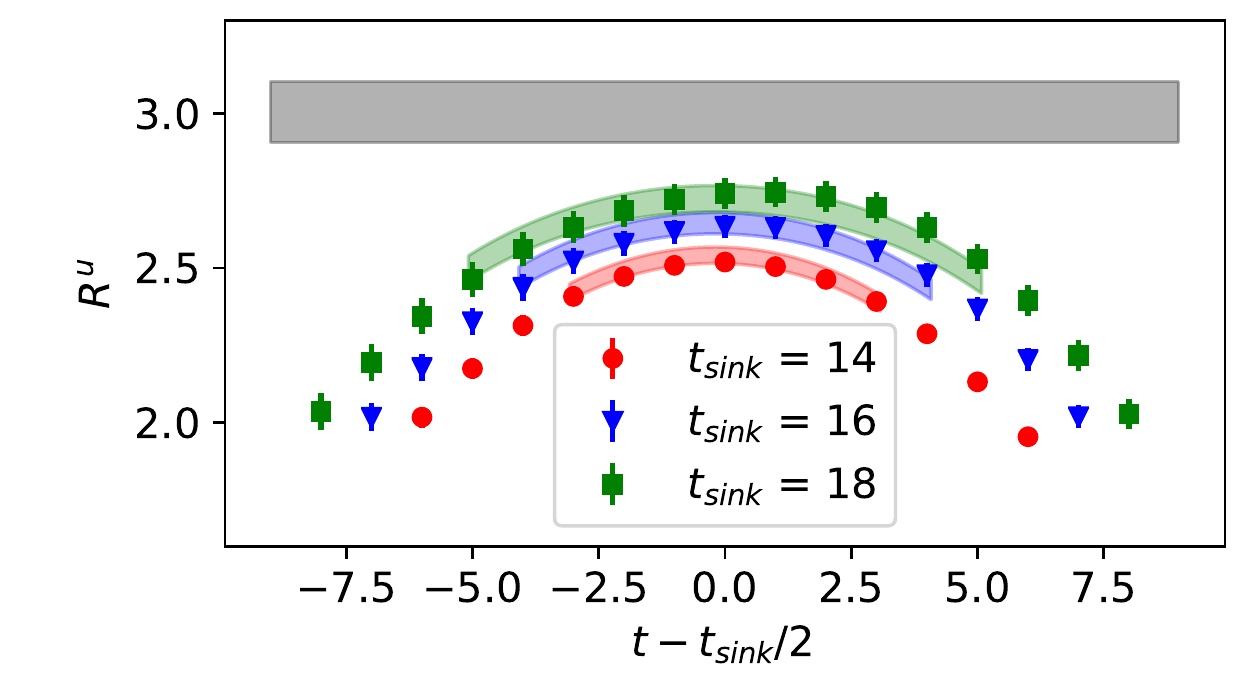}\includegraphics[width =0.44 \textwidth]{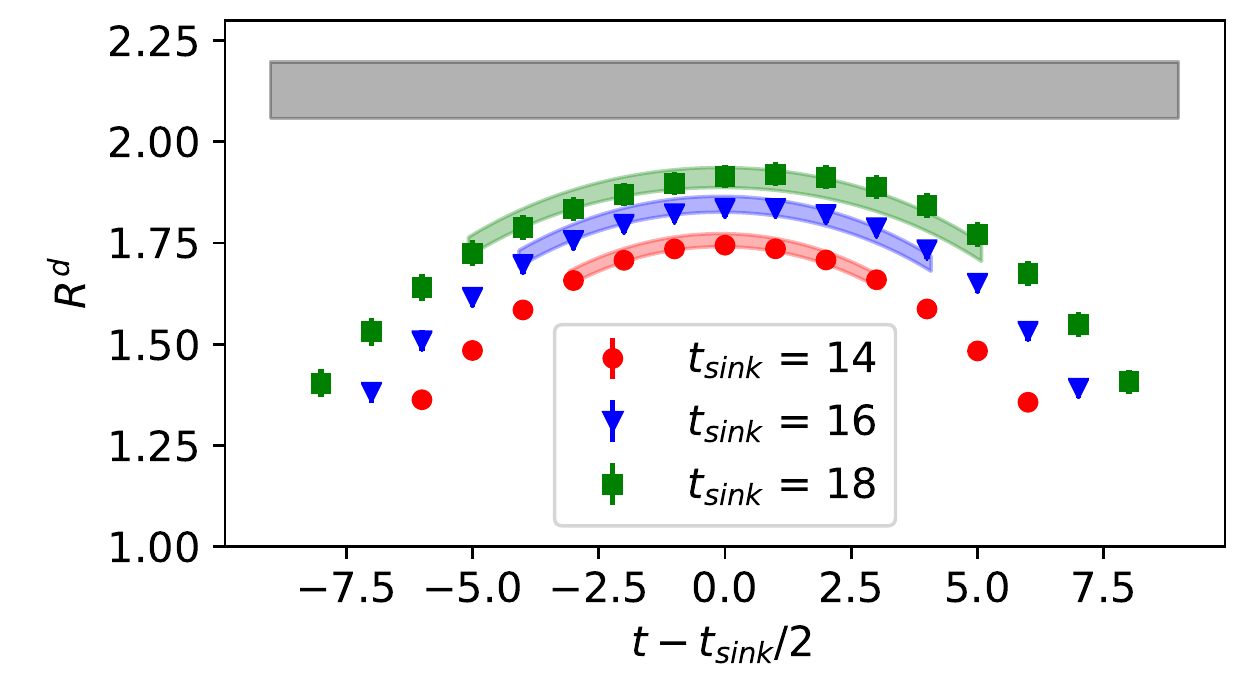}
\includegraphics[width =0.12 \textwidth]{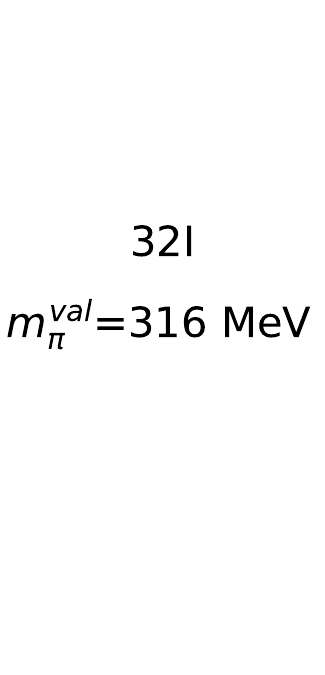}\includegraphics[width =0.44 \textwidth]{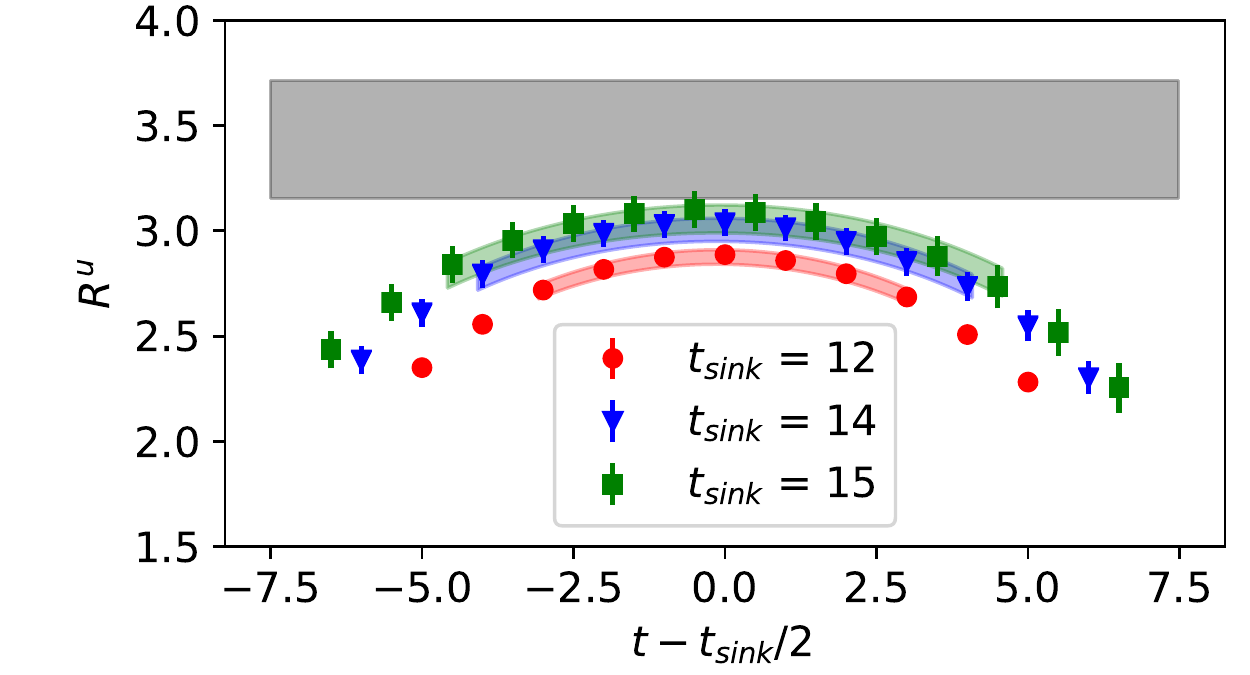}\includegraphics[width =0.44 \textwidth]{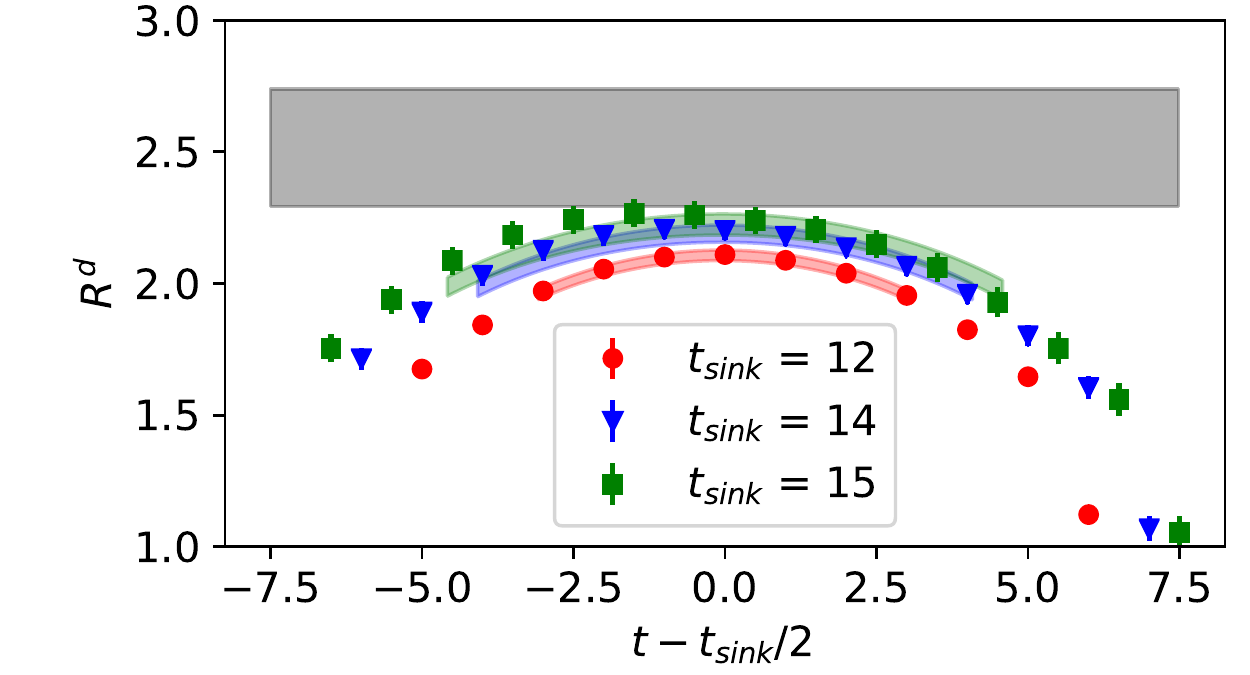}
\includegraphics[width =0.12 \textwidth]{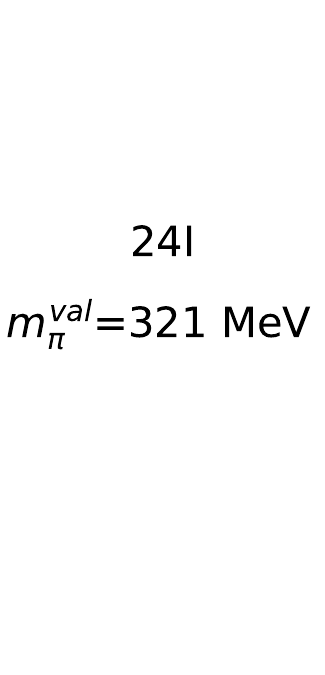}\includegraphics[width =0.44 \textwidth]{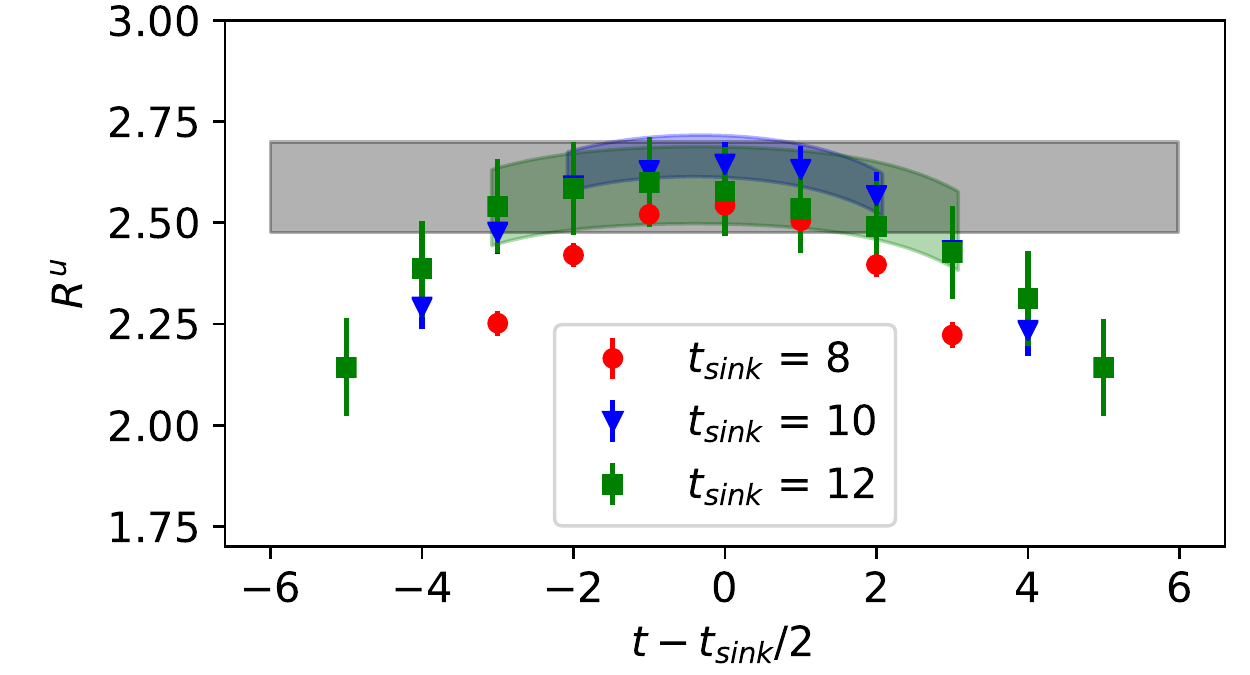}\includegraphics[width =0.44 \textwidth]{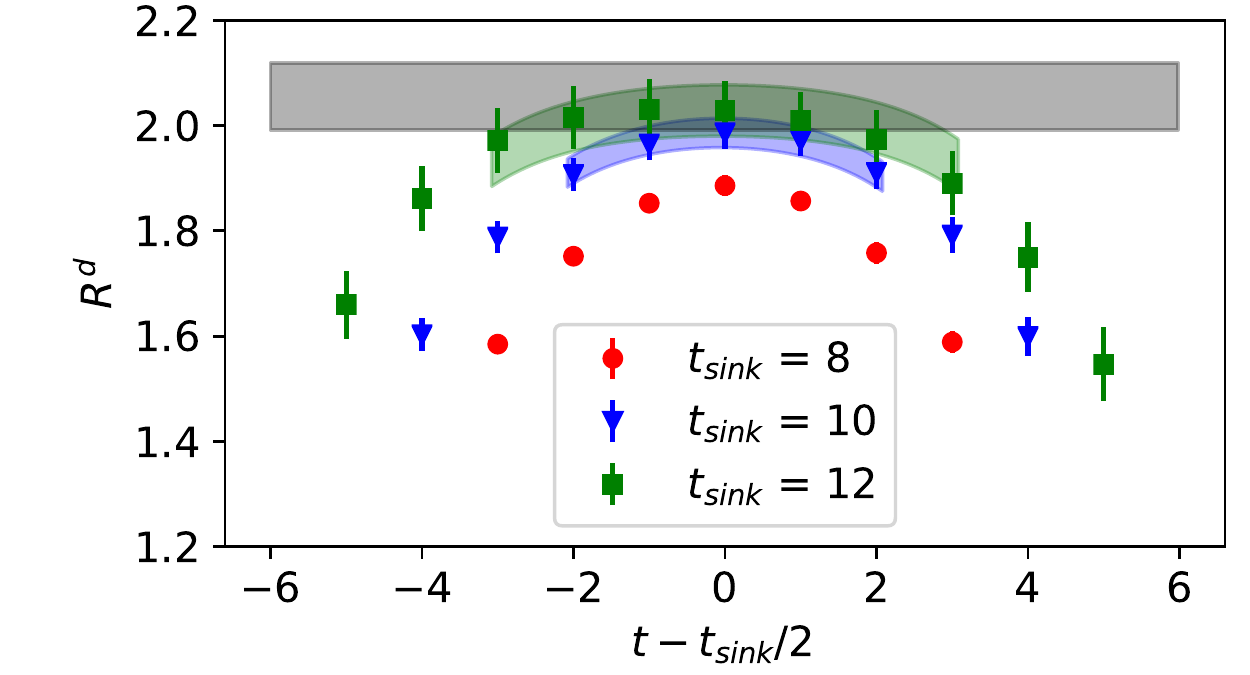}
\includegraphics[width =0.12 \textwidth]{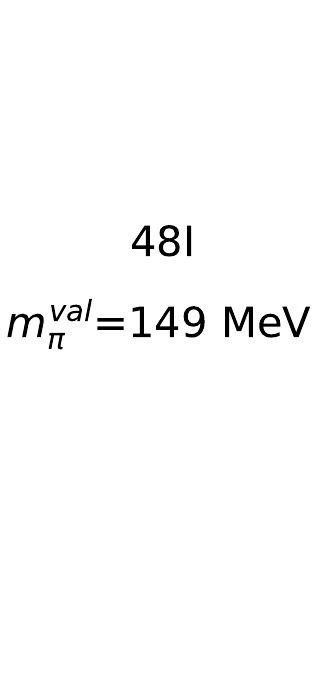}\includegraphics[width =0.44 \textwidth]{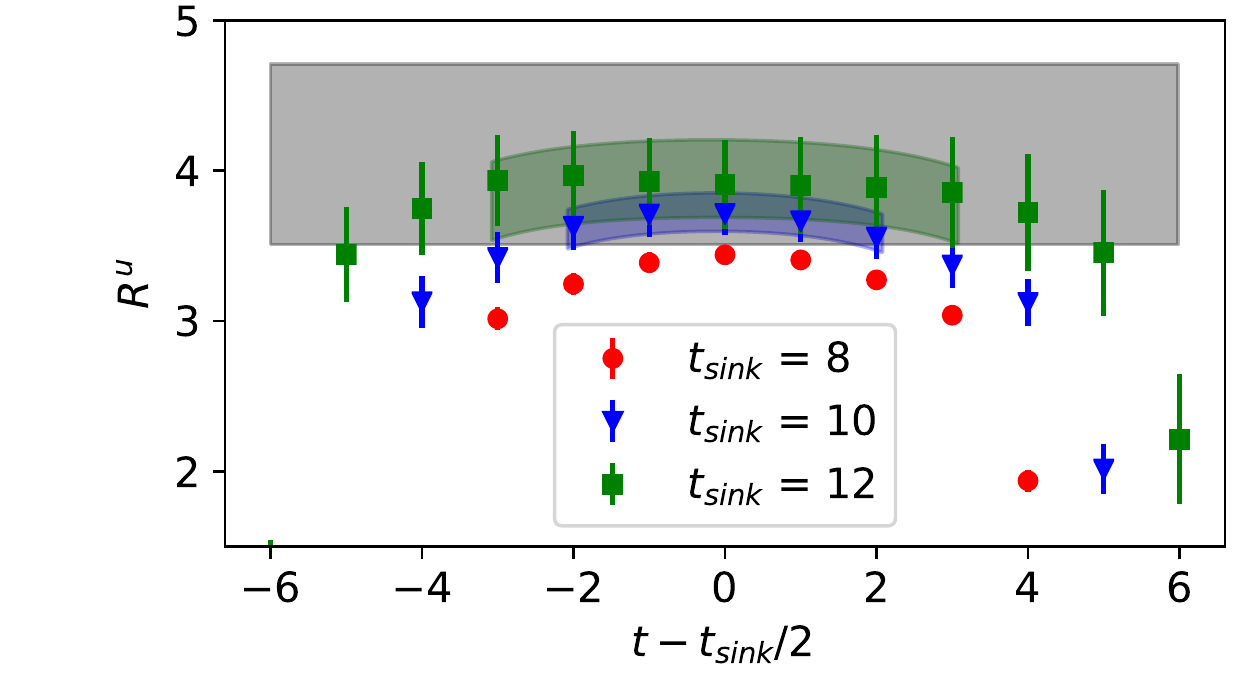}\includegraphics[width =0.44 \textwidth]{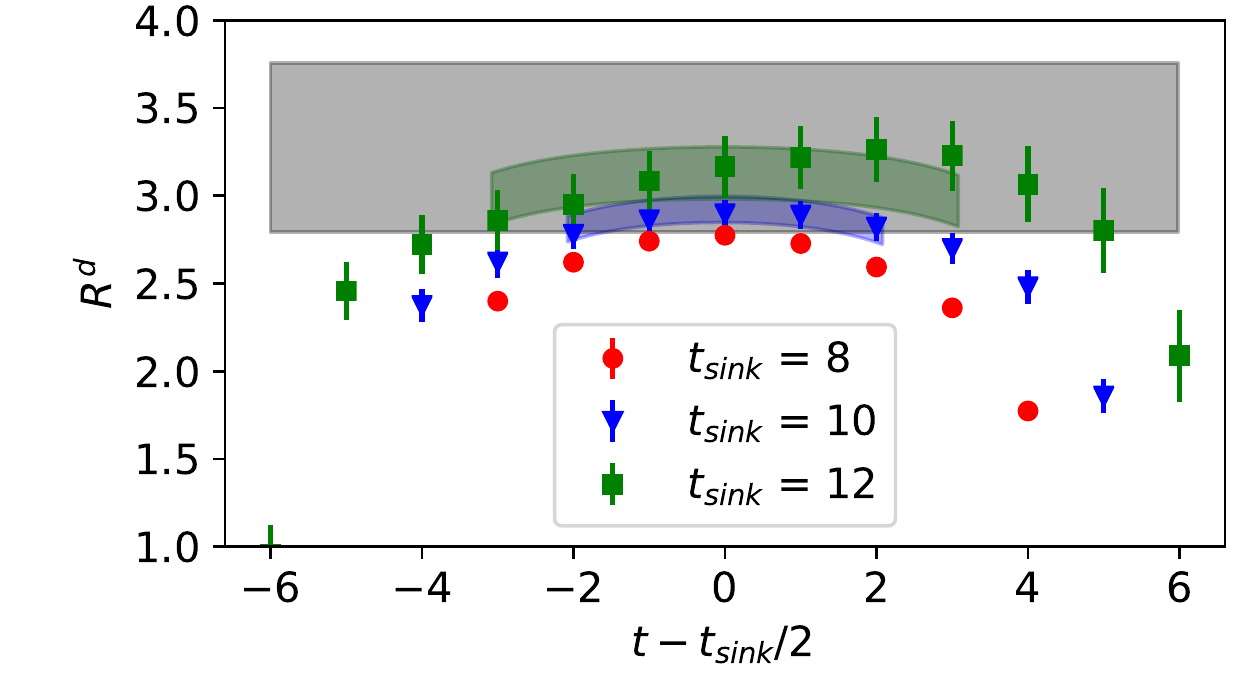}
\includegraphics[width =0.12 \textwidth]{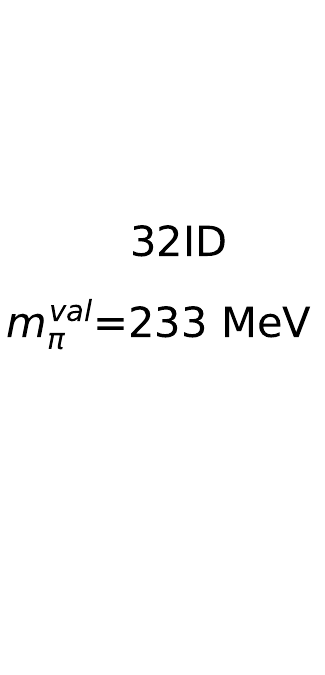}\includegraphics[width =0.44 \textwidth]{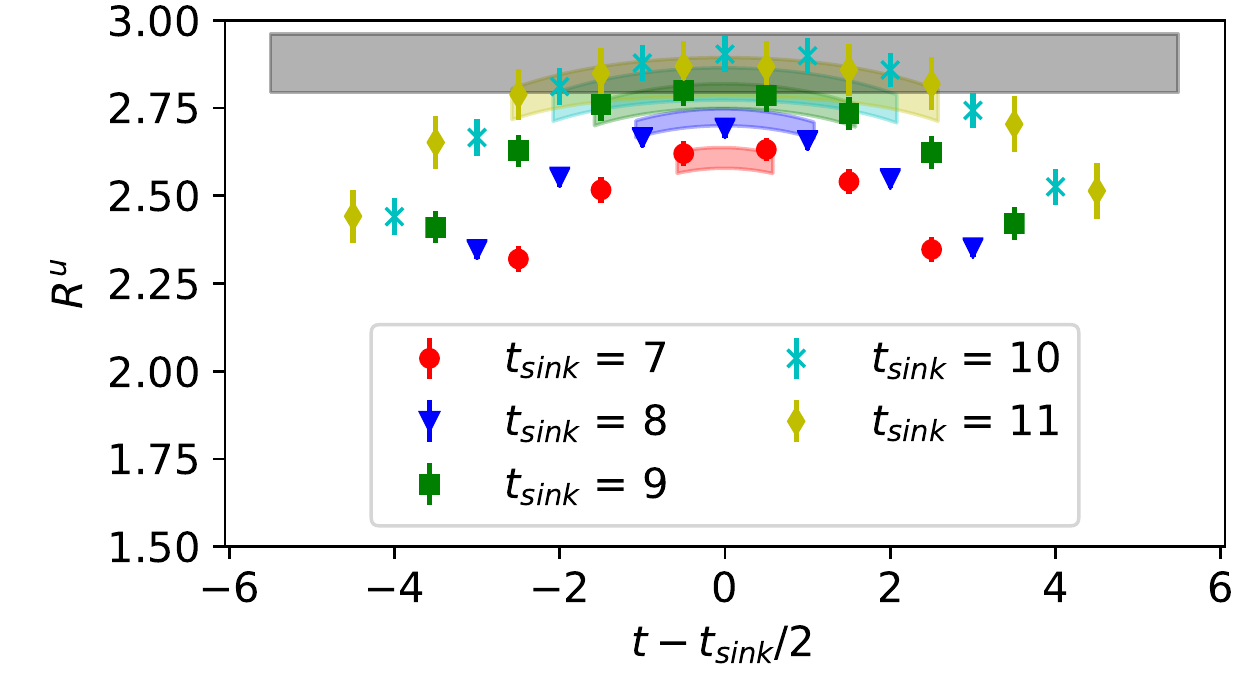}\includegraphics[width =0.44 \textwidth]{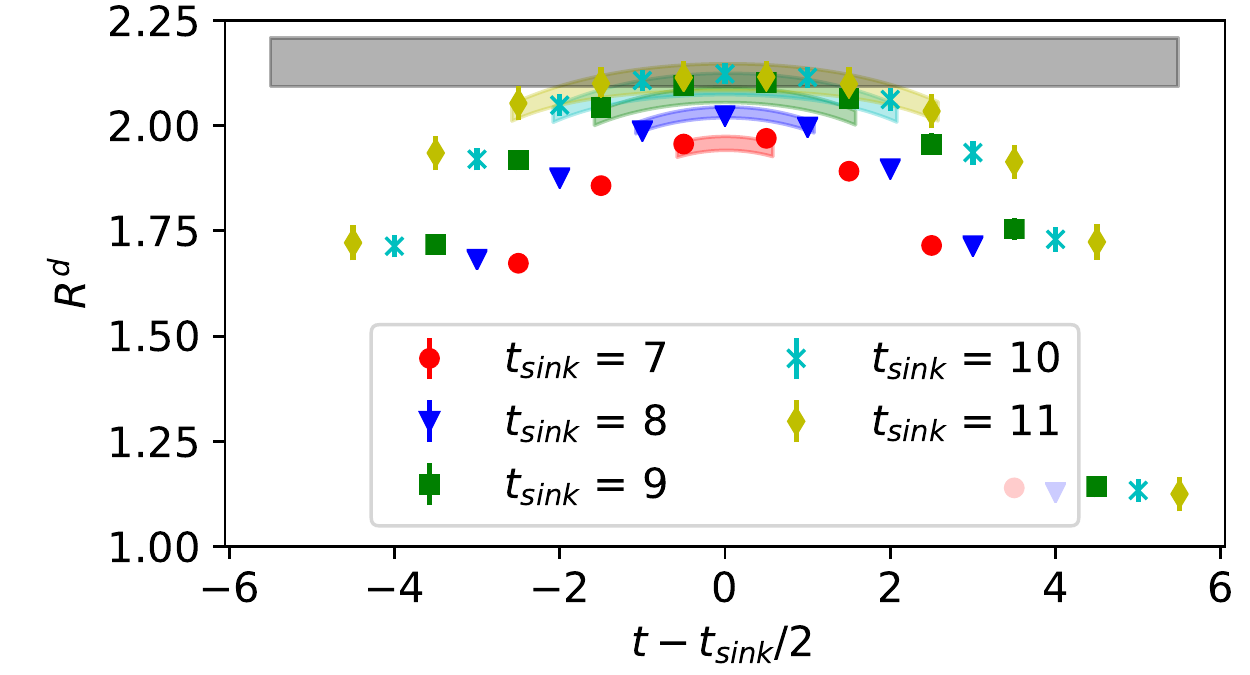}
\caption{$R^u(t_{sink}, t)$(left column) and $R^d(t_{sink}, t)$(right column) as a function of the inserted time $t$ for the five ensembles with the valence pion mass $m_{\pi}^{val}$ as shown. The data points with different source-sink separation $t_{sink}$ are shown in different colors, and the curved bands show the fit to the two-state fit formula Eq.~(\ref{Eq:2statefit}). The constant gray bands show the values of unrenormalized $g_S^u$ and $g_S^d$. The band width indicates one sigma statistical uncertainty. Note that the data points with the smallest $t_{sink}$ value (the red points) are not used in the fit for the ensembles 24I and 48I. }
\label{Fig:2statefit}
\end{figure}

\begin{figure}
\includegraphics[width =0.12 \textwidth]{label_32Ifine_m1.pdf}\includegraphics[width =0.44 \textwidth]{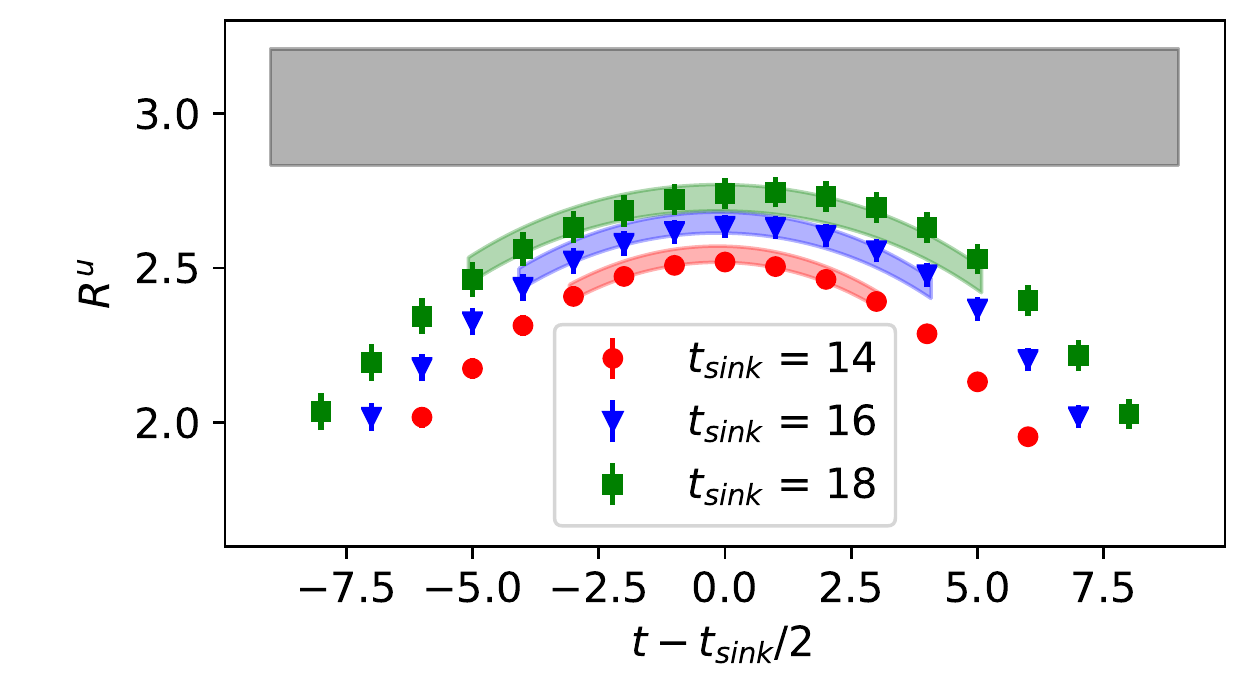}\includegraphics[width =0.44 \textwidth]{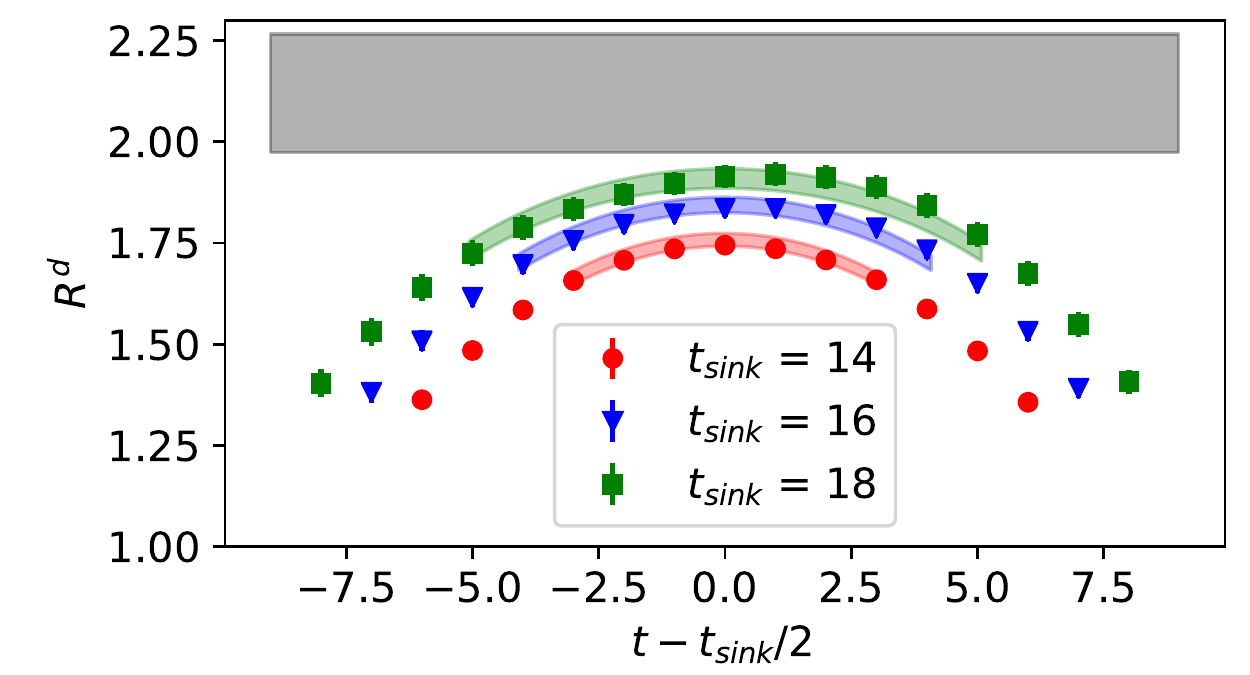}
\includegraphics[width =0.12 \textwidth]{label_32I_m2.pdf}\includegraphics[width =0.44 \textwidth]{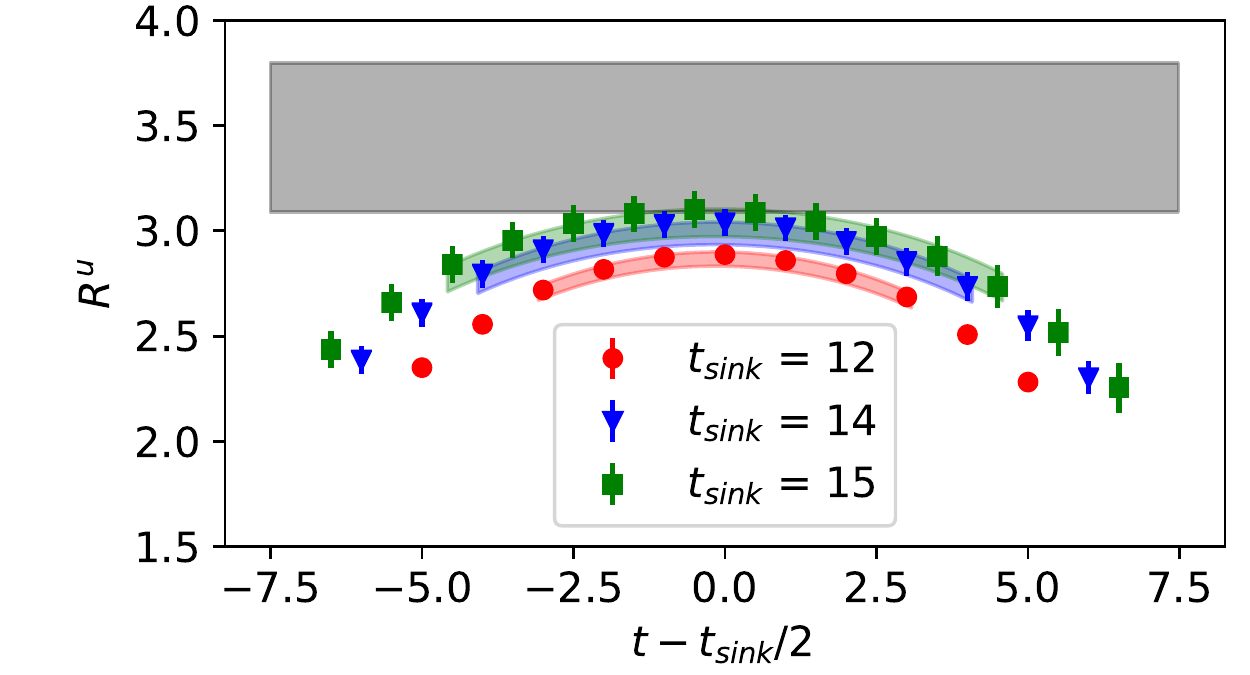}\includegraphics[width =0.44 \textwidth]{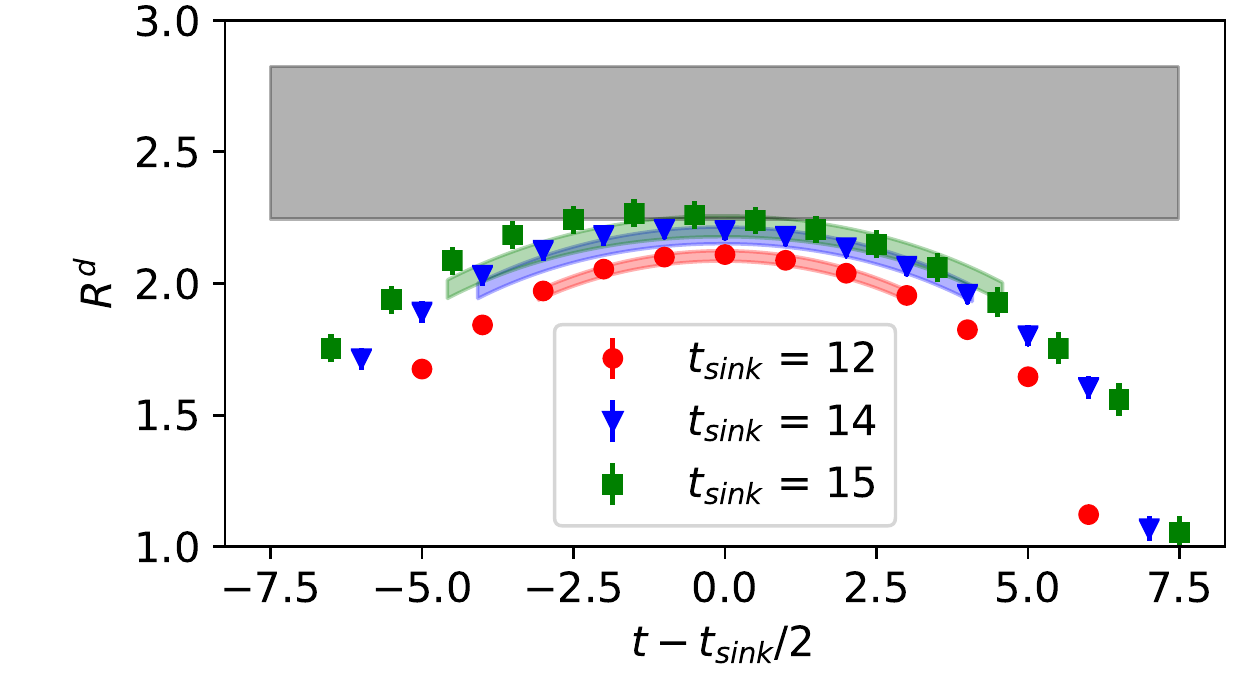}
\includegraphics[width =0.12 \textwidth]{label_24I_m2.pdf}\includegraphics[width =0.44 \textwidth]{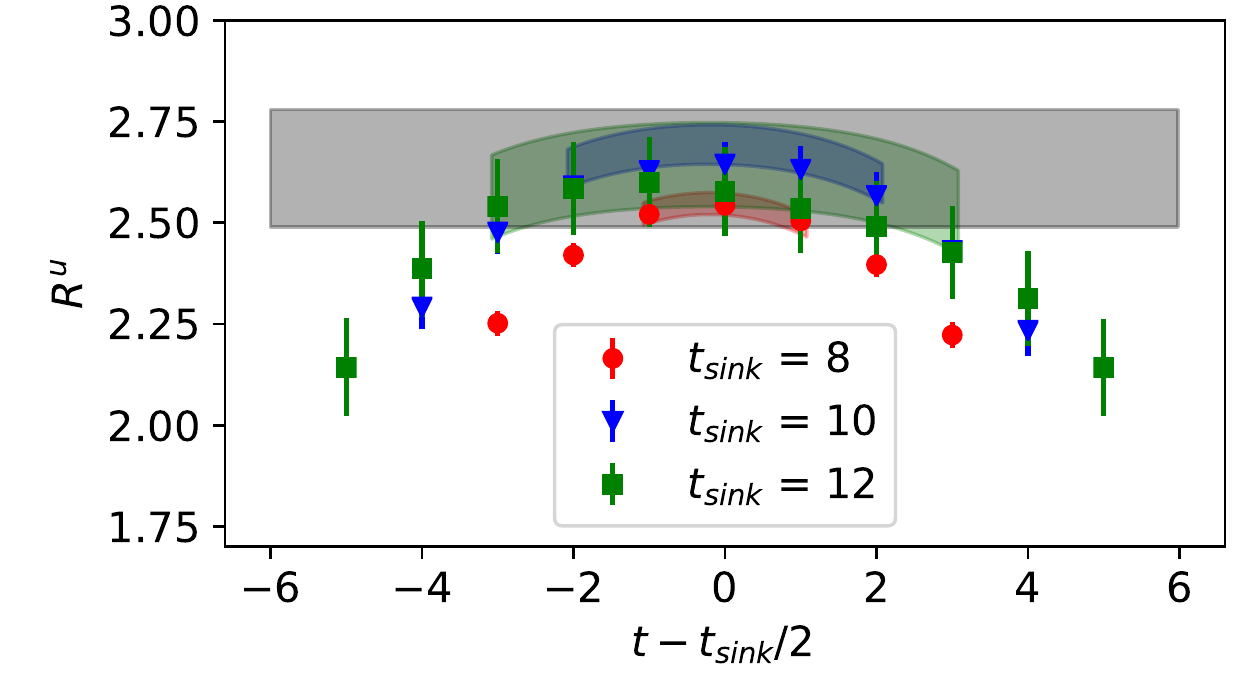}\includegraphics[width =0.44 \textwidth]{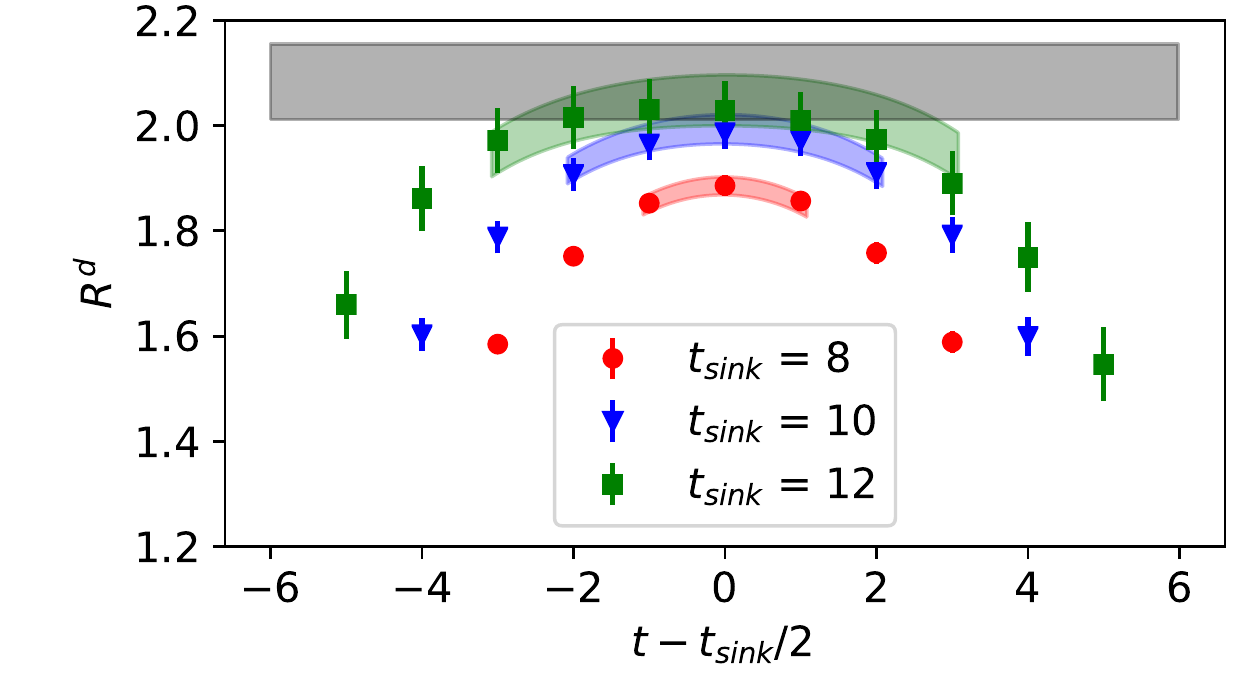}
\includegraphics[width =0.12 \textwidth]{label_48I_m2.pdf}\includegraphics[width =0.44 \textwidth]{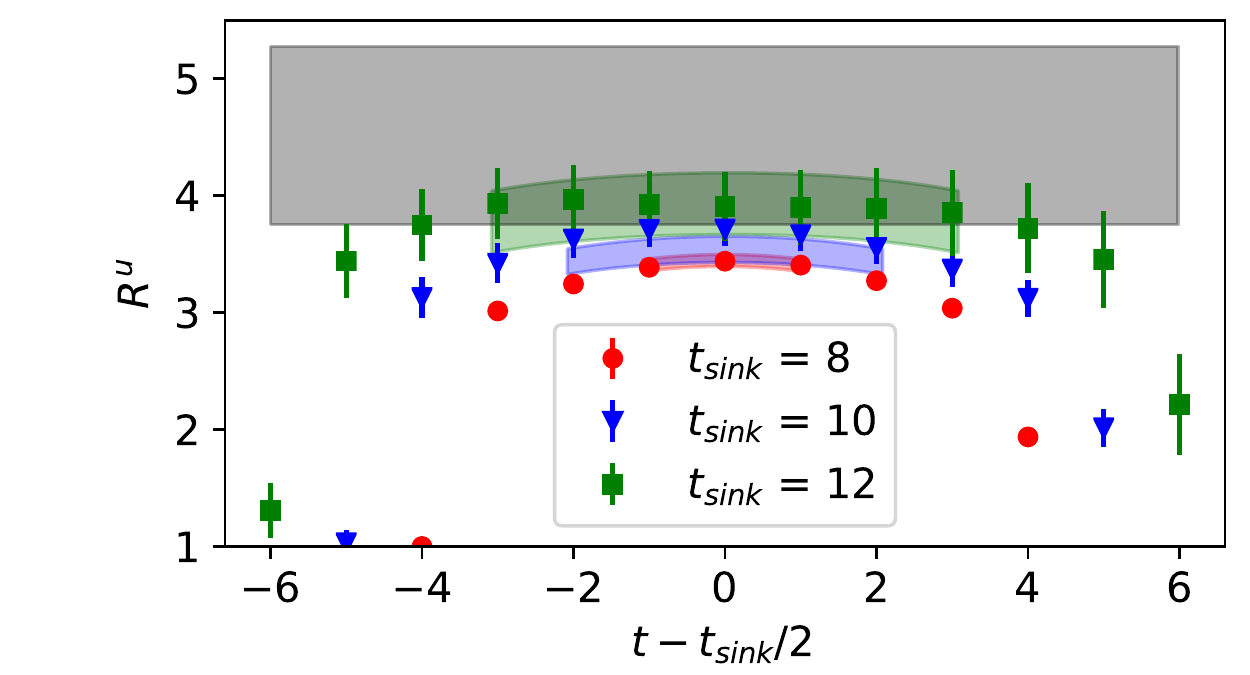}\includegraphics[width =0.44 \textwidth]{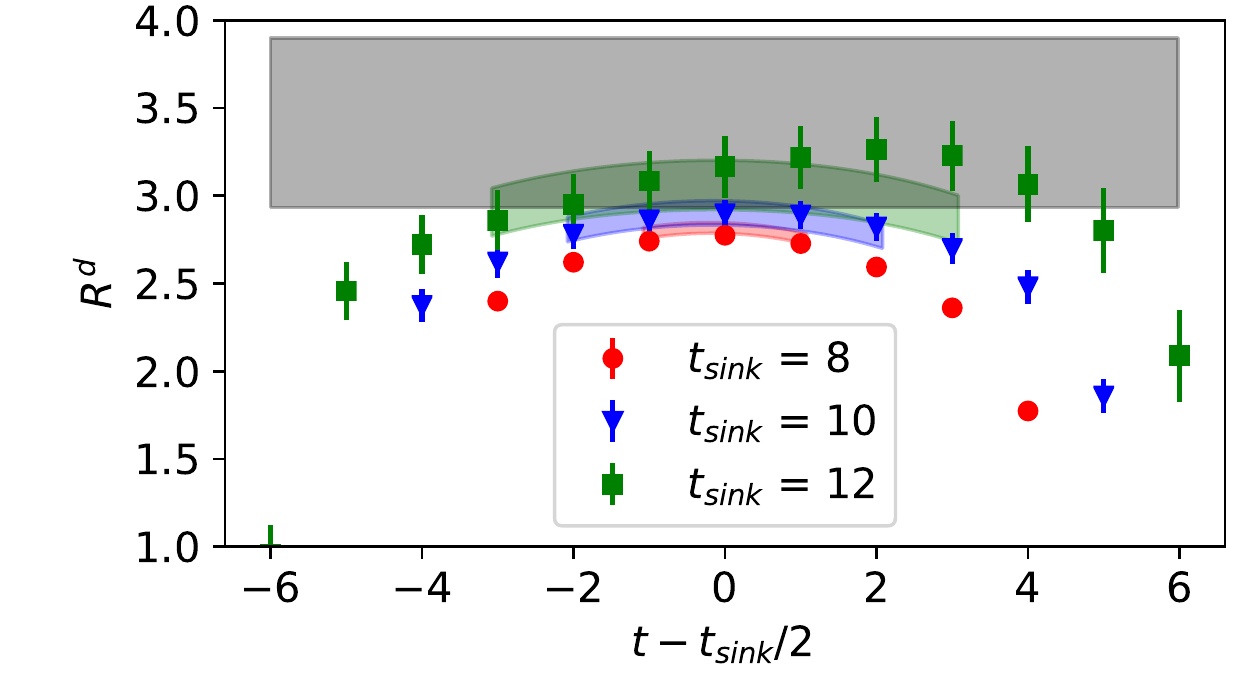}
\includegraphics[width =0.12 \textwidth]{label_32ID_m2.pdf}\includegraphics[width =0.44 \textwidth]{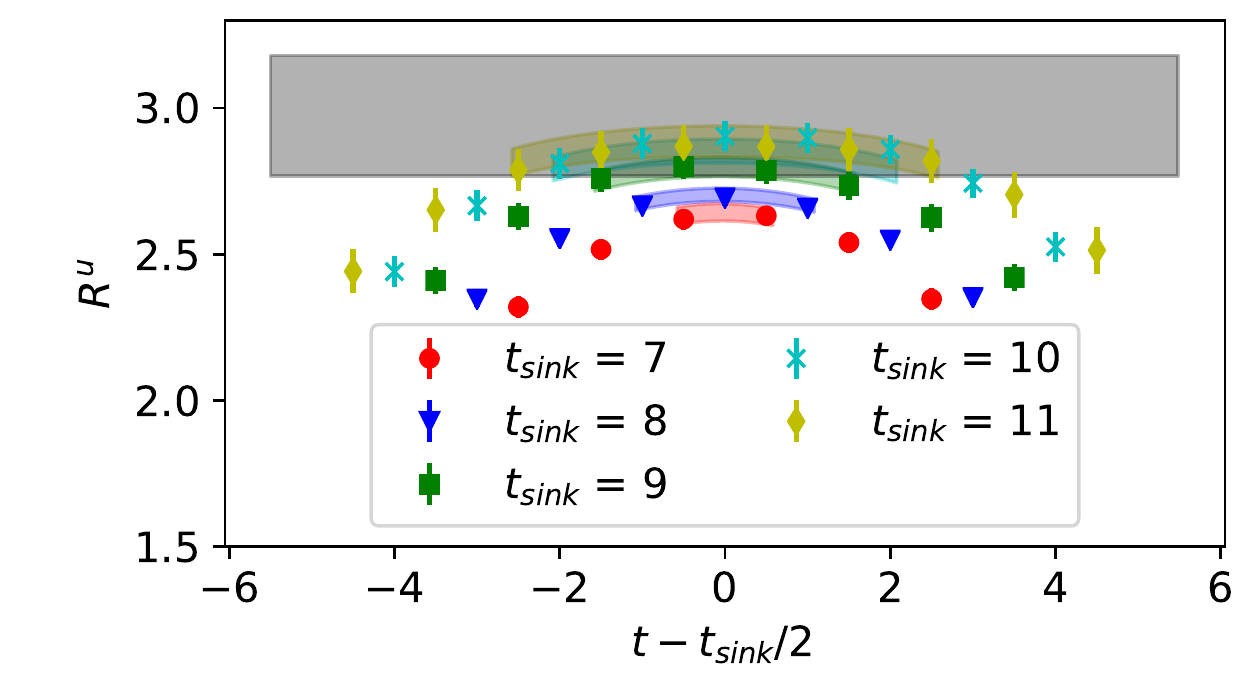}\includegraphics[width =0.44 \textwidth]{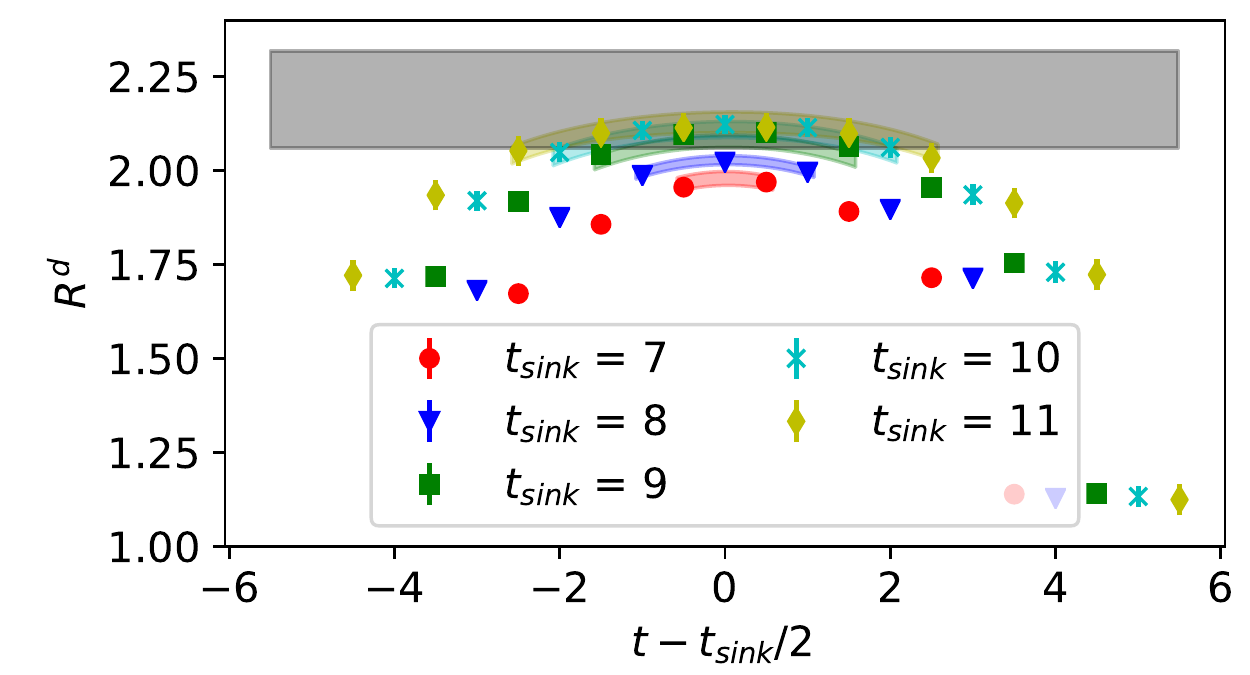}
\caption{$R^u(t_{sink}, t)$(left column) and $R^d(t_{sink}, t)$(right column) as a function of the inserted time $t$ for the five ensembles with the valence pion mass $m_{\pi}^{val}$ as shown. The data points with different source-sink separation $t_{sink}$ are shown in different colors, and the curved bands show the fit to the three-state fit formula Eq.~(\ref{Eq:3statefit}). The constant gray bands show the values of unrenormalized $g_S^u$ and $g_S^d$. The band width indicates one sigma statistical uncertainty. }
\label{Fig:3statefit}
\end{figure}

Fig.~\ref{Fig:2statefit} presents $R^u(t_{sink}, t)$ and $R^d(t_{sink}, t)$ as a function of the insertion time $t$. The data points with different source-sink separation $t_{sink}$ are shown in different colors as indicated in the legend of each plot and the curved bands represent the fit to Eq.~(\ref{Eq:2statefit}). The constant gray bands show the values of unrenormalized $g_S^u$ and $g_S^d$. The width of the bands indicates one-sigma statistical uncertainty. For each ensemble we present one valence pion mass $m_{\pi}^{val}$ as a representative case. Fig.~\ref{Fig:3statefit} is the same as Fig.~\ref{Fig:2statefit} except that the data points are fitted to the three-state fit formula Eq.~(\ref{Eq:3statefit}). 

\begin{table}[t!]
\centering
\begin{tabular*}{.9\textwidth}{@{\extracolsep{\fill}}ccccc}
\hline
\hline
Ensemble ID  & $m_\pi^{sea}$ (MeV)  &$m_\pi^{val}$ (MeV)  &$g_S^{u-d}$(2-state fit)  &$g_S^{u-d}$(3-state fit) \\
\hline 
\multirow{3}{*}{32Ifine} &\multirow{3}{*}{ 371} &344  &0.84(0.06)  &0.86(0.27)\\
                                     &                                 &383 &0.87(0.04) &0.88(0.18)\\
                                     &                                 &420  &0.92(0.04) &0.93(0.21)\\
\hline
\multirow{4}{*}{32I} & \multirow{4}{*}{302}  &295  &0.94(0.15) &0.93(0.21) \\
                               &                                 & 316  &0.91(0.10)  &0.90(0.11)\\
                               &                                 & 353  &0.90(0.06)  &0.91(0.08)\\
                               &                                 & 410  &0.94(0.04)  &0.95(0.04) \\
\hline
\multirow{4}{*}{24I}  & \multirow{4}{*}{337}       &282        &0.48(0.19) &0.47(0.21)  \\
                               &                                        & 321  &0.54(0.09)  &0.53(0.15) \\
                                 &                                       & 348  &0.59(0.07) &0.58(0.12) \\
                                  &                                      & 389  &0.67(0.05) &0.63(0.08) \\                                  
\hline
\multirow{6}{*}{48I} & \multirow{6}{*}{139} & 149  &0.82(0.30)  &1.16(0.56)\\
                               &                                 & 181  &0.73(0.13)  &0.78(0.20)\\
                               &                                 & 207 &0.71(0.08)   &0.73(0.07)\\
                               &                                 &267  &0.91(0.15)   &1.12(0.30)\\
                               &                                 & 331 &0.93(0.08)   &1.14(0.30)\\
                               &                                 &372   &0.97(0.08)  &0.96(0.16)\\
\hline
\multirow{5}{*}{32ID} & \multirow{5}{*}{171} &174 &0.69(0.08)  &0.79(0.99)\\
                                 &                                  &233  &0.73(0.04)  &0.76(0.13)\\
                                 &                                  &262 &0.76(0.03)  &0.76(0.06) \\
                                 &                                  &288 &0.79(0.02)  &0.77(0.04) \\
                                 &                                  &327 &0.82(0.02) & 0.80(0.05)\\
\hline
\hline
\end{tabular*}
\caption{The values of unrenormalized $g_S^{u-d}$ from two- and three-state fits.}
\label{Table:gs_results}
\end{table}

\section{Renormalization}
\label{sec:Renormalization}

\begin{table}[t!]
\begin{tabular*}{.9\textwidth}{@{\extracolsep{\fill}}cccccc}
\hline
\hline
 &32Ifine &32I &24I  &48I &32ID\\
 \hline
$Z_S$ &0.951(2)(14) &1.018(1)(15) &1.117(1)(16)  &1.135(1)(16) &1.236(1)(20) \\
\hline   
\hline
\end{tabular*}
\caption{The renormalization factor for all ensembles}
\label{table:ZS}
\end{table}

We use the regularization independent momentum subtraction (RI/MOM) scheme~\cite{Martinelli:1994ty,Bi:2017ybi} under the Landau gauge to renormalize the scalar quark bilinear operator. The quark self energy is defined through the axial vector normalization constant $Z_A$ following the definition of the RI/MOM scheme~\cite{Bi:2017ybi}. Thanks to the statistical enhancement of using the volume source propagator~\cite{Chen:2017mzz}, the statistical uncertainty at a given RI/MOM scale $p^2$ can be smaller than 0.1\% and the major uncertainty of $Z_S$ comes from the systematic one of the estimated 4-loop effect in the perturbative matching between the RI/MOM and $\overline{\textrm{MS}}$ schemes, and also the value of $\Lambda_{QCD}$, scale running, lattice spacing, and fit range. The overall uncertainty on most of the ensembles is about 1.5\%~\cite{Liang:2021pql}, but it is slightly larger on the 32ID ensemble since the minimum $p^2$ we used is smaller and thus the matching uncertainty is larger.

\begin{figure}
\includegraphics[width =0.7\textwidth]{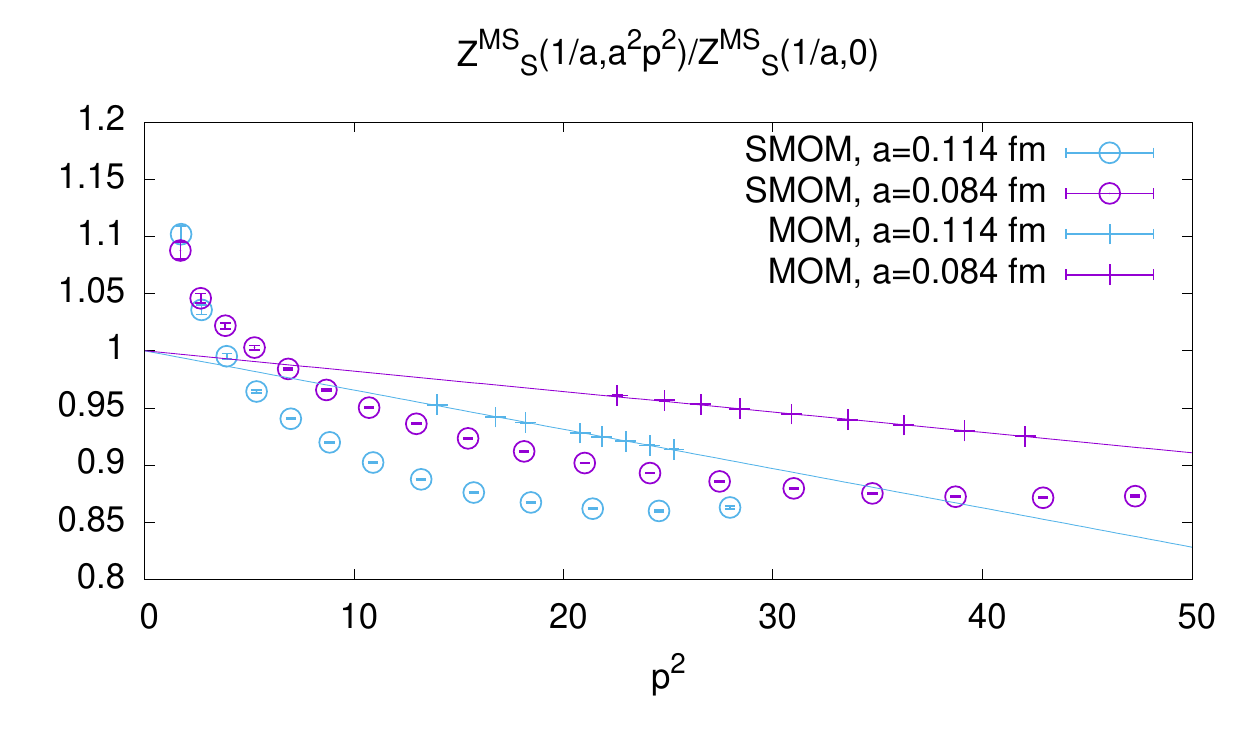}
\caption{Comparing $Z^{\overline{\textrm{MS}}}_S(2 \textrm{GeV})$ values using different RI/MOM and RI/SMOM scales, at two lattice spacings. The values are normalized with the  $a^2p^2$ extrapolated values using the RI/MOM scheme to show  the non-perturbative and discretization effects}
\label{Fig:compare_smom}
\end{figure}

We also investigated the RI/SMOM scheme~\cite{Aoki:2007xm,Sturm:2009kb} which has a better perturbative matching convergence at least up to 2-loop level~\cite{Sturm:2009kb}. Based on the calculation on the 64I ensemble which has the same setup as the physical point ensemble 48I but smaller lattice spacing (0.084 fm) and also that on 48I (0.114 fm), we obtained the $Z^{\overline{\textrm{MS}}}_S(2 \textrm{GeV})$ values using different RI/MOM and RI/SMOM scales, as shown in Fig.~\ref{Fig:compare_smom}. The RI/MOM cases use the momenta along the body diagonal direction with cutoff $\sum_{\mu}p_{\mu}^4/(\sum_{\mu}p_{\mu}^2)^2<0.26$ (blue and purple crosses for the results on the coarser and finer lattice spacings), and the RI/SMOM cases use the surface-diagonal momenta such as $p=(k,k,0,0)$ and $p'=(k,0,k,0)$ (blue and purple dots). All the data are normalized by the $a^2p^2$ extrapolated values of $Z^{\overline{\textrm{MS}}}_S(2 \textrm{GeV})$ using the RI/MOM scheme, to show a comparison of the non-perturbative and discretization effects. From the figure, we can see that the non-linear behavior of $Z_S$ in the $p^2 \in$[4-10] GeV$^2$ region observed in the  48I case (blue dots, the same as we obtained in Ref.~\cite{Bi:2017ybi}) using the RI/SMOM scheme remains in the 64I case (purple dots), and thus is not a discretization effect. At the same time, the non-linear $p^2$ dependence for $p^2 $ $>$ 10 GeV$^2$ becomes weaker at smaller lattice spacing, and thus it would be a discretization effect at ${\cal O}(a^4p^4)$. Generally speaking, the $p^2$ dependence of the SMOM case is highly non-linear and it is hard to find a linear window to eliminate the discretization error. On the other hand, the result through the MOM scheme shows perfect linear $p^2$ dependence with the slope decreasing in $a^2$. Thus we use the RI/MOM scheme in this work and leave further comparisons between MOM and SMOM schemes to a separate work.

Our results of the scalar current renormalization constants are listed in Table~\ref{table:ZS}, with two uncertainties from the statistics and systematics.

\section{Results}
\label{sec:Results}
\begin{figure}
\includegraphics[width =0.5 \textwidth]{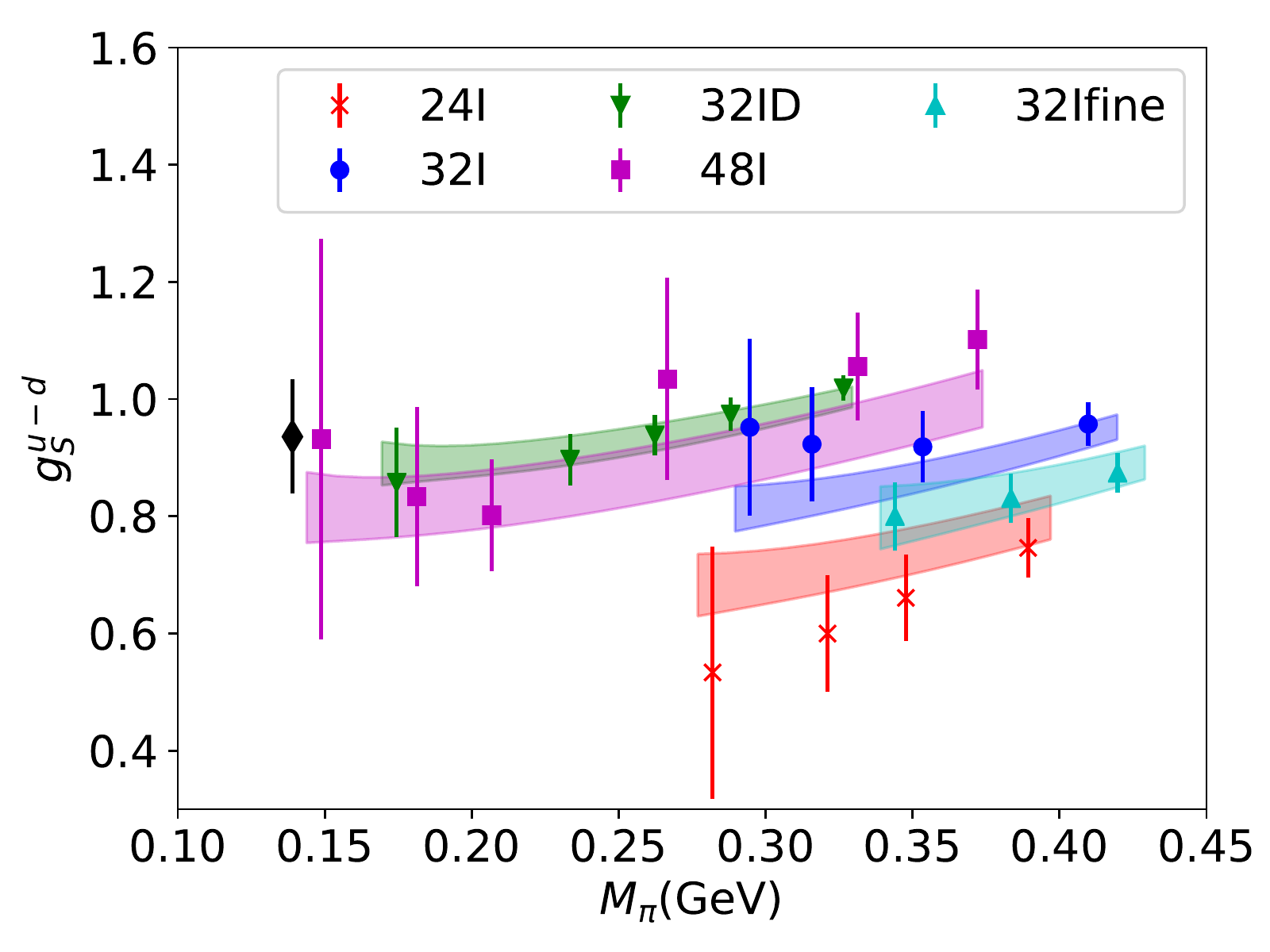}\includegraphics[width =0.5 \textwidth]{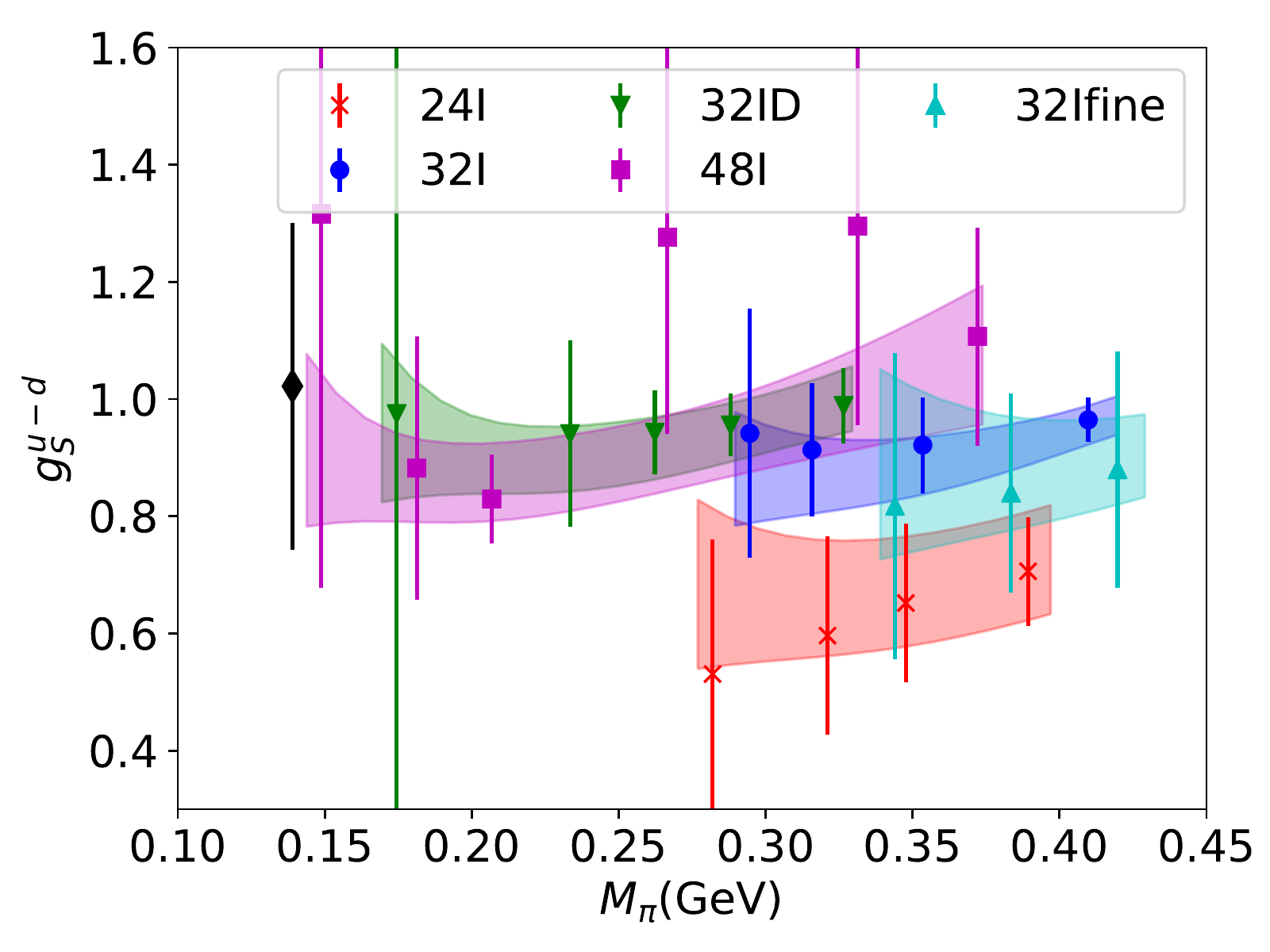}
\caption{$g_S^{u-d}$ as a function of valence pion mass. The left(right) panel shows the values of $g_S^{u-d}$ extracted from the two(three)-state fits. The curved bands show the joint fit of all data points to the formula in Eq.~(\ref{Eq:extrapolation}). The width of the bands indicates one-sigma statistical error. The black diamond is the extrapolated value of $g_S^{u-d}$ at the physical pion mass, continuum limit and infinite volume.  }
\label{Fig:extrapolation}
\end{figure}

The renormalized values of $g_S^{u-d}$ extracted from the two(three)-state fit for all ensembles  are shown in the left(right) panel of Fig.~\ref{Fig:extrapolation} as a function of valence pion mass. In order to obtain the result at the physical point, we perform a joint fit of all data points to the following form, 
\begin{equation}
g_S^{u-d}(m_{\pi, val}^2, m_{\pi, sea}^2, a, L) = C_0 + C_1 m_{\pi, val}^2 + C_2 m_{\pi, sea}^2 + C_3 a^2 + C_4 e^{-m_{\pi, val} L}.
\label{Eq:extrapolation}
\end{equation} 
Notice that the gauge action of the ensemble 32ID is different from that of the other ensembles; the coefficient of the $a^2$ term for 32ID should not be the same with the others. We denote it by $C_3^\prime$. The fitted parameters and the value of $\chi^2$/d.o.f. are listed in Table~\ref{Table:extrapolation}. In Fig.~\ref{Fig:extrapolation}, the curved bands show the fit to Eq.~\ref{Eq:extrapolation} and the black diamond indicates the value of $g_S^{u-d}$ extrapolated to the physical pion mass, continuum limit and infinite volume limit. The extrapolated value at the physical point is $0.94 \pm 0.10$ for the two-state fit case and $1.0 \pm 0.3$ for the three-state fit case. We take the result of two-state fit as our final result and the difference between the central values of two- and three-state fits as an estimation of the systematic uncertainty due to the excited-states contamination. 

In order to investigate the systematics in the extrapolation, we performed the following alternative fits. 1) We performed the extrapolation with different formulas by adding a log term $m_{\pi}^2 \log m_{\pi}^2$ and using a different volume dependent term $\frac{m_\pi^2}{\sqrt{m_\pi L}} e^{-m_\pi L}$, and the extrapolated results did not change. 2) For each ensemble, there are 1-2 valences pion masses that are very close to the sea pion mass, i.e., $m_\pi^{val}$ = 383MeV for 32Ifine, $m_\pi^{val}$ = 295MeV and 316MeV for 32I, $m_\pi^{val}$ = 321MeV and 348MeV for 24I, $m_\pi^{val}$ = 149MeV for 48I and  $m_\pi^{val}$ = 174MeV for 32ID.   We fit these 7 data points to the formula $g_S^{u-d}(m_{\pi, val}^2, m_{\pi, sea}^2, a) = C_0 + C_1 (m_{\pi, val}^2 + m_{\pi, sea}^2) + C_3 a^2$. The volume dependent term is ignored since we found this term is not important in the fit.  The extrapolated value of $g_S^{u-d}$ is 0.93(0.21). The agreement between the results of this fit using the nearly unitary data points and the original fit using all data points supports the validity of our partially quenched scheme. 3) We dropped the data point of 32Ifine in fit 2) since it has the largest pion mass, and redid the fit. The extrapolated result is  $g_S^{u-d}$ = 0.99(0.21). The systematic error in the extrapolation is then estimated by the differences between the original fit and these alternative fits. 

Another source of systematics comes from the uncertainties of the renormalization factors, which are $\sim 1.5\%$ as shown in Table~\ref{table:ZS}. This systematic error is estimated by $1.5\%$ of the central value of $g_{S}^{u-d}$ and is added quadratically to the systematic error due to the excited-states contamination and the extrapolation to obtain the total systematic error. 

Our final result of the isovector scalar charge is
\begin{equation}
g_{S}^{u-d}= 0.94 (10)_{stat}(8)_{sys}
\end{equation}
where the first error is the statistical error and the second error is the systematic error. 

\begin{table}[t!]
\centering
\begin{tabular*}{.9\textwidth}{@{\extracolsep{\fill}}cccccccc}
\hline
\hline
& $C_0$         &$C_1(\textrm{GeV}^{-2})$        &$C_2(\textrm{GeV}^{-2})$          &$C_3(\textrm{fm}^{-2})$          &$C_3^\prime(\textrm{fm}^{-2})$  &$C_4$       &$\chi^2$/d.o.f. \\
\hline
two-state fit & 0.95(0.11)   &1.8(0.3)  &-2.7(0.7)   &-12(6)  &-3(4)      &2(2)  &0.68 \\
three-state fit & 1.06(0.29)   &2.5(1.6)  &-4.7(2.1)   &-19(17)  &-10(10)      &8(10)  &0.19 \\
\hline
\hline
\end{tabular*}
\caption{Results of fitting the $g_S^{u-d}$ values extracted from the two- and three-state fits to the formula in Eq.~(\ref{Eq:extrapolation}) . }
\label{Table:extrapolation}
\end{table} 

\begin{figure}
\includegraphics[width =0.6 \textwidth]{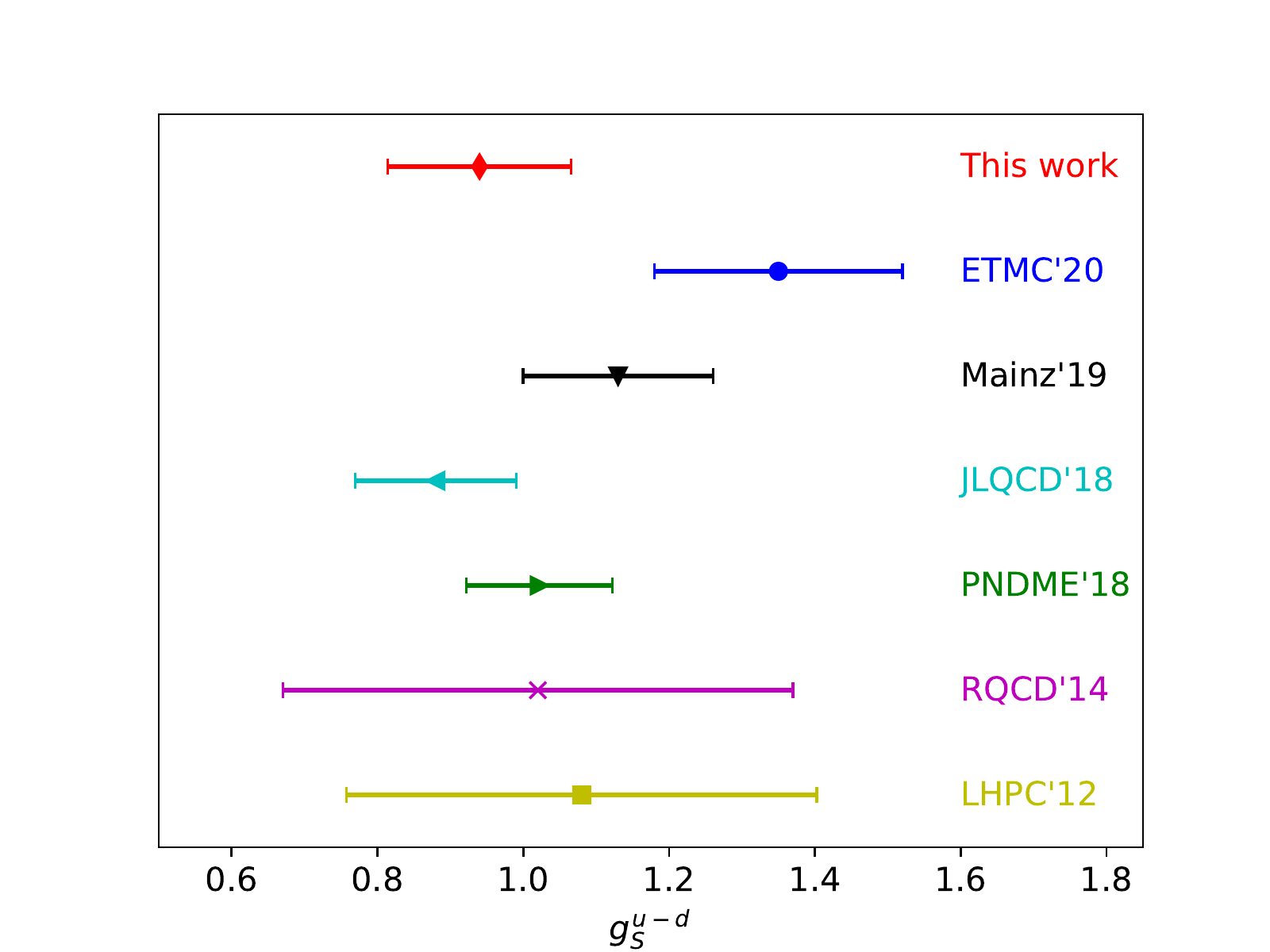}
\caption{Extrapolation.}
\label{Fig:gS_compare}
\caption{Comparison of our result (red diamond) for $g_S^{u-d}$ with a number of other lattice QCD calculationss: Mainz'19 ~\cite{Harris:2019bih}, JLQCD'18 ~\cite{Yamanaka:2018uud}, PNDME'18 ~\cite{Gupta:2018qil}, ETMC'20~\cite{Alexandrou:2019brg},  RQCD'14~\cite{Bali:2014nma} and LHPC'12 ~\cite{Green:2012ej}. The error bars are the statistical and systematic errors added quadratically.}
\end{figure}

In Fig.~\ref{Fig:gS_compare}, we compare our result with a number of other lattice calculations: ETMC'20~\cite{Alexandrou:2019brg}, Mainz'19~\cite{Harris:2019bih}, JLQCD'18~\cite{Yamanaka:2018uud}, PNDME'18~\cite{Gupta:2018qil}, RQCD'14~\cite{Bali:2014nma} and LHPC'12~\cite{Green:2012ej}. ETMC'20~\cite{Alexandrou:2019brg} presented the results from a $N_f=2+1+1$ twisted mass ensemble with physical pion mass and lattice spacing $a\sim 0.08$ fm. Mainz'19~\cite{Harris:2019bih} computed the isovector scalar charge on a set of $N_f=2+1$ ensembles with improved Wilson fermions, covering four values of lattice spacing and pion mass range 200 - 350 MeV. The chiral, continuum and finite-size extrapolations have been performed. JLQCD'18~\cite{Yamanaka:2018uud} performed the calculations using dynamical overlap fermions with four pion masses in the range 290 - 540 MeV and a single lattice spacing $a\sim 0.11$ fm. The PNDME'18~\cite{Gupta:2018qil} results were obtained from mixed-action calculations using clover valence action on HISQ ensembles at four lattice spacings and three pion masses in the range 135--320 MeV. RQCD'14~\cite{Bali:2014nma} obtained the results from $N_f=2$ clover ensembles at three lattice spacings and several pion masses with the lowest value at 150 MeV. LHPC'12 has analyzed a number of Wilson clover and Domain-wall ensembles as well as a mixed-action scheme which uses a Domain-wall action on staggered sea quarks. We compare only the latest results from each group. The previous work from the ETMC~\cite{Alexandrou:2017qyt} and the PNDME~\cite{Bhattacharya:2016zcn, Bhattacharya:2013ehc} collaborations are not included in the comparison. 

As shown in Refs.~\cite{Liang:2018pis,Liu:2020okp}, there is no
mixing from the glue and quark disconnected insertions to the connected insertions. Thus, it is meaningful to define $u$ and $d$ contributions  separately in the connected insertions. Such separation can be compared to those from the DIS and Drell-Yan experiments for $\langle x\rangle$ and $g_A$ for example, when such separation is accommodated in the global analysis of the parton distribution functions~\cite{Liu:2020okp}. Here we present the extrapolated values of the scalar charges for $u$ and $d$ from the connected insertions:
\begin{equation}
g_S^u(\textrm{CI}) = 4.04(19)_{stat}(54)_{sys}, \,\,\, g_S^d(\textrm{CI}) = 3.11(14)_{stat}(64)_{sys},
\end{equation}
where the systematic errors are estimated by the same method as for $g_S^{u-d}$. By the same token, the isoscalar matrix element for the connected insertion in the $\overline{\rm{MS}}$ scheme at 2 GeV is
\begin{equation}
 g_S^{u+d}(\textrm{CI})= g_S^u(\textrm{CI}) + g_S^d(\textrm{CI}) =7.18(32)_{stat}(80)_{sys}. 
 \end{equation}
The systematic uncertainties of $g_S^u(\textrm{CI})$, $g_S^d(\textrm{CI})$ and $g_S^{u+d}(\textrm{CI})$ mainly come from the alternative fit 2) as described above. There is a large cancellation in this source of systematic uncertainty when taking the difference of $g_S^u(\textrm{CI})$ and $g_S^d(\textrm{CI})$, therefore it dose not contribute much to the systematic uncertainty of $g_S^{u-d}$. 

\section{Summary}
\label{sec:Summary}
We have presented the result of the nucleon scalar charge from a lattice calculation using overlap fermions on domain-wall configurations. The calculation is performed on five ensembles with various values of the pion mass, lattice spacing and volume, covering the pion mass from the physical value to 371 MeV and lattice spacing from 0.06 fm to 0.14 fm. Using the multi-mass algorithm for overlap fermions, 3--6 valence quark masses are obtained for each ensemble. Extrapolation to the physical point is obtained by fitting all the data points to a formula originated from partially quenched chiral perturbation theory. To control the excited-states contamination, the correlation functions are computed at several values of source-sink time separation that are around or larger than 1 fm. We performed two- and three-state fits, and the results generally agree with each other within statistical errors. The differences between them are taken as a systematic error due to the excited-states contamination. 
Our final result of the nucleon isovector scalar charge is $g_{S}^{u-d}= 0.94 (10)_{stat}(8)_{sys}$ in the $\overline{\rm{MS}}$ scheme at 2 GeV.

\section*{Acknowledgements}
We thank the RBC/UKQCD Collaboration for sharing their domain-wall gauge configurations with us.
LL thanks the support from the Strategic Priority Research Program of Chinese Academy of Sciences with Grant No.\ XDB34030301, the CAS Interdisciplinary Innovation Team program and Guangdong Provincial Key Laboratory of Nuclear Science with No.\ 2019B121203010. 
JL thanks the support from the Guangdong Major Project of Basic and Applied Basic Research No.\ 2020B0301030008 and Science and Technology Program of Guangzhou No.\ 2019050001.
This work is partially supported by the U.S. DOE Grant DE-SC0013065 and DOE Grant No.\ DE-AC05-06OR23177 which is within the framework of the TMD Topical Collaboration.
This research used resources of the Oak Ridge Leadership Computing Facility at the Oak Ridge National Laboratory, which is supported by the Office of Science of the U.S. Department of Energy under Contract No.\ DE-AC05-00OR22725. This work used Stampede time under the Extreme Science and Engineering Discovery Environment (XSEDE), which is supported by National Science Foundation Grant No. ACI-1053575.
We also thank the National Energy Research Scientific Computing Center (NERSC) for providing HPC resources that have contributed to the research results reported within this paper.
We acknowledge the facilities of the USQCD Collaboration used for this research in part, which are funded by the Office of Science of the U.S. Department of Energy.

\bibliographystyle{apsrev}
\bibliography{reference}

\end{document}